\documentclass[11pt]{article}
\usepackage{amsmath,amssymb,amsthm}
\usepackage{geometry}
\usepackage{hyperref}
\usepackage{graphicx}
\usepackage{booktabs}
\usepackage{float}
\usepackage{algorithm}
\usepackage{algpseudocode}

\graphicspath{{./figures/}}

\newtheorem{theorem}{Theorem}[section]
\newtheorem{lemma}[theorem]{Lemma}
\newtheorem{proposition}[theorem]{Proposition}
\newtheorem{corollary}[theorem]{Corollary}

\title{Representation-Aware Closed-Surface Parameterization: Exact Sensitivity Analysis and Pole-Aware Planar Preconditioning}
\author{Zixi Yi$^{1}$, Gary P. T. Choi$^{2}$\\
\\
\footnotesize{$^{1}$School of Mathematics and Statistics, Central South University}\\
\footnotesize{$^{2}$Department of Mathematics, The Chinese University of Hong Kong}
}
\date{ }

\begin{document}
\maketitle

\begin{abstract}
Surface parameterization is a fundamental task in geometry processing and shape analysis. In particular, for topologically closed surfaces, it is common to parameterize them onto a standardized closed domain in $\mathbb{R}^3$, with a prime example being a unit sphere for genus-0 closed surfaces. For simplicity, many parameterization methods use planar projections and perform computations on the plane rather than directly on the 3D surface. However, perturbations introduced before projection or geometric evolution propagate through the computation and may affect map quality, convergence, stopping behavior, and geometric validity. To address this issue, in this work we formulate representation sensitivity as a branchwise input--output problem, separating smooth propagation through active Jacobians from changes in anchors, stopping rules, and fallback events. Under consistent piecewise-linear assembly, the unmodified south-pole Beltrami correction in the FLASH method for spherical conformal parameterization reduces exactly to source-mesh harmonic extension. On a fixed-anchor branch, the corrected map factors into chart transition, harmonic propagation, and inverse stereographic projection, yielding an explicit sensitivity Jacobian and a finite-triangle conditioning bound. Separately, we introduce a bounded five-parameter affine preconditioner over planar scale, translation, and anisotropy and evaluate candidates on the corrected spherical map. Across 30 genus-0 surfaces, it reduces median polar upper-tail log-condition and chord-angle distortion by $10.17\%$ and $10.97\%$, respectively, with both measures improving on every mesh. The harmonic realization agrees with the original FLASH correction to numerical precision and gives a median correction-stage speedup of $1.63\times$. We apply the same framework to other spherical mapping methods, including the conformalized mean-curvature flow (cMCF) and spherical density-equalizing maps (SDEM), to quantify transient amplification, terminal attenuation, and changes in stopping behavior and geometric validity, and the experimental results show fewer residual flips and increasingly consistent improvements in fixed-template remeshing. Altogether, our work highlights the importance of representation selection as a structural and practical component of spherical parameterization and downstream computation.
\end{abstract}

\section{Introduction}
\label{sec:introduction}

Parameterization of closed surfaces provides a common computational domain for surface registration, shape comparison, remeshing, and related imaging and geometry-processing tasks~\cite{ShefferPraunRose2006}. In particular, for genus-0 closed surfaces, it is common to consider spherical parameterizations. Existing approaches include spherical conformal parameterizations~\cite{haker2000conformal,GuYau2003,GotsmanGuSheffer2003,GuWangChanThompsonYau2004,KazhdanSolomonBenChen2012,ChoiLamLui2015FLASH,LiaoHuangLinYueh2024}, spherical area-preserving parameterizations~\cite{HuangLiaoLin2024Equiareal,LyuLuiChoi2024SDEM,LiuYueh2026SAEM}, and methods that balance angle and area distortion~\cite{NadeemSuZengKaufmanGu2017}. Other constructions promote local rigidity or use hierarchical optimization~\cite{WangLiuLiu2014ARAP,HuFuLiu2018AHSP}, control quasi-conformal distortion or enforce injectivity~\cite{ChoiManLui2016QC,SchmidtBornCampenKobbelt2019}, or support remeshing and surface sampling~\cite{PraunHoppe2003,ChoiHoLui2016SpherePointCloud}. 

It is noteworthy that directly solving for the optimal parameterizations on the spherical domain is usually computationally challenging. For instance, the harmonic energy minimization approach~\cite{GuWangChanThompsonYau2004} involves solving a nonlinear optimization problem on the sphere via an iterative scheme. Also, the conformalized mean-curvature flow (cMCF) method~\cite{KazhdanSolomonBenChen2012} evolves the surface through geometry-dependent implicit updates, and the spherical density-equalizing map (SDEM) method~\cite{LyuLuiChoi2024SDEM} deforms a spherical map iteratively based on the physical principle of density diffusion with certain projection steps involved. To simplify the computation, many prior methods have considered the use of planar charts or intermediate embeddings to represent the closed target surface. For instance, Haker et al.~\cite{haker2000conformal} developed a linearized method for computing spherical conformal parameterizations via solving the Laplace equation on a planar domain followed by the stereographic projection onto the sphere. The FLASH method~\cite{ChoiLamLui2015FLASH} further combined planar computations with quasi-conformal pole correction based on the linear Beltrami solver (LBS)~\cite{LuiLamWongGu2013LBS} to improve conformality near the pole. 




While surface parameterization methods are typically evaluated through map-level quantities such as angle and area distortion and local injectivity, numerical accuracy in parameterization-based computations also depends on how geometry is represented in the parameter domain. Aflalo et al.~\cite{AflaloKimmelZibulevsky2013} showed that controlling the conformal factor is important for accurate geodesic computation on a flattened domain. In stereographic constructions, pole-related distortion has also motivated the use of curved parameter elements~\cite{ChenXiaoChen2025}. These observations suggest treating representation changes as an input--output sensitivity problem, in the sense of classical conditioning analysis~\cite{Higham2002}. In FLASH's stereographic pipeline~\cite{ChoiLamLui2015FLASH}, chart perturbations change the finite target geometry and enter the subsequent pole correction. In iterative methods, initial-state perturbations enter geometry-dependent updates and can affect the trajectory, stopping index, and geometric validity. On a fixed execution path consisting of differentiable operations, the chain rule propagates local sensitivities through the active computations~\cite{GriewankWalther2008}. A representation-aware analysis must therefore account for both smooth propagation within a fixed branch and changes in discrete choices such as anchors, stopping rules, and fallback events.

\begin{figure}[t!]
\centering
\includegraphics[width=\linewidth]
{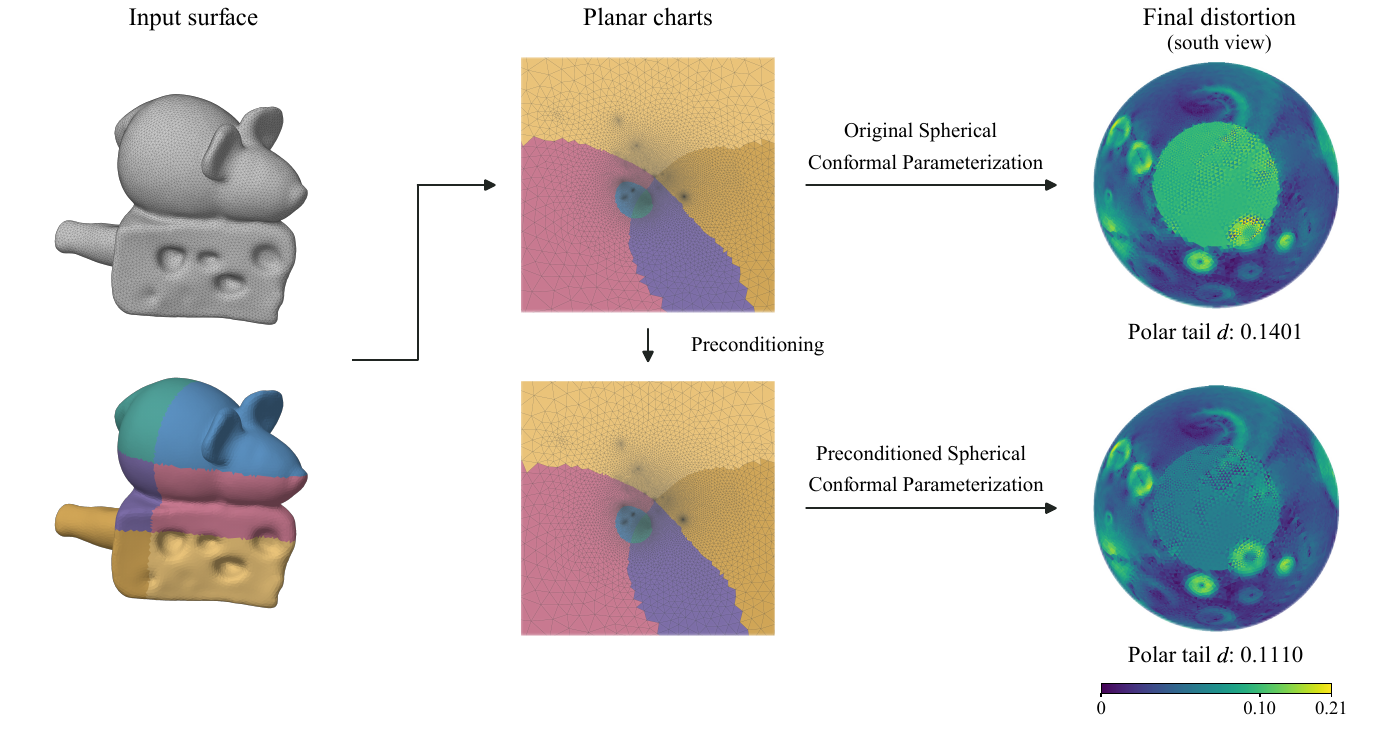}
\caption{
\textbf{An illustration of the pole-aware planar preconditioning for spherical parameterization of genus-0 closed surfaces.} The left panel shows the input genus-0 Mouse surface model with a mesh view (top) and an alternative visualization with different regions in different colors. The upper row of the right panel shows the original pipeline of the FLASH method~\cite{ChoiLamLui2015FLASH} for spherical conformal parameterization, while the lower row shows the pipeline with the proposed bounded five-parameter affine planar preconditioner. The planar charts illustrate the representation adjustment, and the final corrected spherical maps are viewed from the south pole and colored by facewise log-condition distortion $d_T=\log\kappa_2(J_T)$. The polar upper-tail log-condition score decreases by $20.80\%$ after preconditioning.
}
\label{fig:preconditioning-mouse}
\end{figure}

In this work, we formulate representation sensitivity in spherical parameterization as a branchwise input--output problem. For the projection-based FLASH method for spherical conformal parameterization, we derive an exact reduction of the unmodified south-pole Beltrami correction to harmonic extension with the intrinsic source-mesh Dirichlet operator (see Figure~\ref{fig:preconditioning-mouse} for an illustration). On a fixed-anchor branch, this reduction gives a closed-form factorization of the corrected map and an explicit Jacobian that separates chart transition, harmonic propagation, and inverse stereographic projection. We also derive a finite-triangle conditioning bound that relates the effect of inverse stereographic projection to planar element shape and radial variation. This factorization identifies how planar representation changes reach the final spherical map and provides the basis for a bounded five-parameter affine preconditioner over planar scale, translation, and anisotropy. Candidate planar representations are generated through a multistart search subject to north--south size-balance constraints. Their quality is assessed on the fully corrected FLASH maps, so the optimization targets the representation ultimately passed to downstream computation while preserving distortion and geometric-validity constraints. We further apply the representation-sensitivity framework to other representative spherical parameterization methods, including cMCF and SDEM, as mechanism-level studies of initial-state perturbation propagation. Products of state-dependent step Jacobians reveal transient amplification, terminal attenuation, and changes in stopping behavior and geometric validity. These experiments quantify how representation perturbations are transmitted through iterative parameterization algorithms and identify trajectory-level quantities that can guide future sensitivity-aware initialization. For FLASH, the proposed preconditioning reduces the median polar upper-tail log-condition and chord-angle distortions by $10.17\%$ and $10.97\%$, respectively, across a cohort of 30 genus-0 closed surfaces, with both measures improving on every mesh. Downstream experiments further show that preconditioned FLASH initializations reduce residual flips in SDEM and improve local conditioning and element quality in fixed-template remeshing. The main contributions of this work are as follows:

\begin{itemize}
\item We formulate representation sensitivity as a branchwise input--output problem for closed-surface parameterization, with applications to genus-0 spherical parameterization algorithms.

\item For the FLASH method for spherical conformal parameterization, we prove that, under consistent piecewise-linear assembly and the unmodified Beltrami correction, the LBS stiffness matrix coincides with the intrinsic Dirichlet stiffness matrix of the source mesh. This identity yields an exact source-mesh harmonic extension, a fixed-anchor closed-form representation of the corrected map, and a factorized sensitivity Jacobian. Numerical tests validate the Jacobian and verify the equivalence of the original and harmonic correction realizations.

\item We apply the framework to cMCF and SDEM to quantify how initial-state perturbations propagate through state-dependent iterations, including transient amplification, terminal attenuation, stopping-index changes, and loss of geometric validity.

\item We derive a finite-triangle conditioning bound and combine it with the exact FLASH sensitivity structure to design a bounded five-parameter affine preconditioner. Experiments evaluate its effect on the corrected spherical map and its downstream benefits for SDEM and fixed-template remeshing.
\end{itemize}

The remainder of the paper is organized as follows. Section~\ref{sec:rs-framework} introduces the proposed representation-perturbation framework and the fixed-branch stability condition. Section~\ref{sec:flash-exact} develops the exact chart dependence of the FLASH correction, including the harmonic reduction, the fixed-anchor factorization, and the pole-aware Jacobian. Section \ref{sec:other-realizations} applies the perturbation framework to cMCF and SDEM. Section~\ref{sec:preconditioning} presents the bounded affine preconditioner and the constrained search procedure. Section \ref{sec:experiment-design} reports the numerical validation and downstream experiments. We discuss the principal findings and limitations in Section~\ref{sec:discussion} and conclude our work in Section~\ref{sec:conclusion}.

\section{Discrete setting and representation perturbations}
\label{sec:rs-framework}

For the method-specific analysis and numerical experiments below, let $\mathcal M=(V,F)$ be a connected, closed, oriented, nondegenerate genus-0 triangular mesh with $n$ vertices. A spherical realization $S\in(\mathbb S^2)^n$ assigns a unit vector to each vertex and retains the source connectivity. Following standard elementwise analysis of triangulated surface parameterizations~\cite{FloaterHormann2005}, we measure distortion on the affine map from each source face to its target chord triangle, the flat triangle joining the three spherical vertices. Spherical areas are used separately for area diagnostics and validity checks. This distinction prevents smooth conformality, chord-triangle distortion, and spherical area distortion from being conflated.

A representation perturbation changes the numerical input while the algorithmic settings remain fixed. For the FLASH method for spherical conformal parameterization~\cite{ChoiLamLui2015FLASH}, it acts on the planar coordinates after north--south balancing and on the initial state in the iterative experiments. The resulting trajectories are compared under the same mesh, solver, and stopping rules.

\subsection{Propagation on a fixed computational branch}
Let $X_0$ be a baseline state and write $X^{(0)}_\theta=G_\theta(X_0)$, $X^{(k+1)}_\theta=T_{k,\sigma_k}(X^{(k)}_\theta)$, and $S_\theta=\Psi(X^{(K)}_\theta)$, where $G_0$ is the identity. The discrete variable $\sigma_k$ records choices such as anchors, retries, and fallback rules. We call a neighborhood a fixed branch when these choices and the stopping index $K$ remain constant. For differentiable active maps, the chain rule gives
\begin{align}
 \mathrm D_\theta S_\theta
 &=\mathcal P_\theta\,\mathrm D_\theta G_\theta(X_0),
 \label{eq:rs-chain}\\
 \mathcal P_\theta
 &=\mathrm D\Psi(X^{(K)}_\theta)
   \mathrm DT_{K-1,\sigma_{K-1}}(X^{(K-1)}_\theta)
   \cdots\mathrm DT_{0,\sigma_0}(X^{(0)}_\theta).
 \label{eq:rs-propagator}
\end{align}
The rightmost factor selects the perturbation directions; $\mathcal P_\theta$ propagates them through the solver. Analytic active maps yield an analytic composition on the same branch. These are standard composition properties, used here to organize method-specific calculations.

\begin{figure}[t!]
\centering
\includegraphics[width=\linewidth]{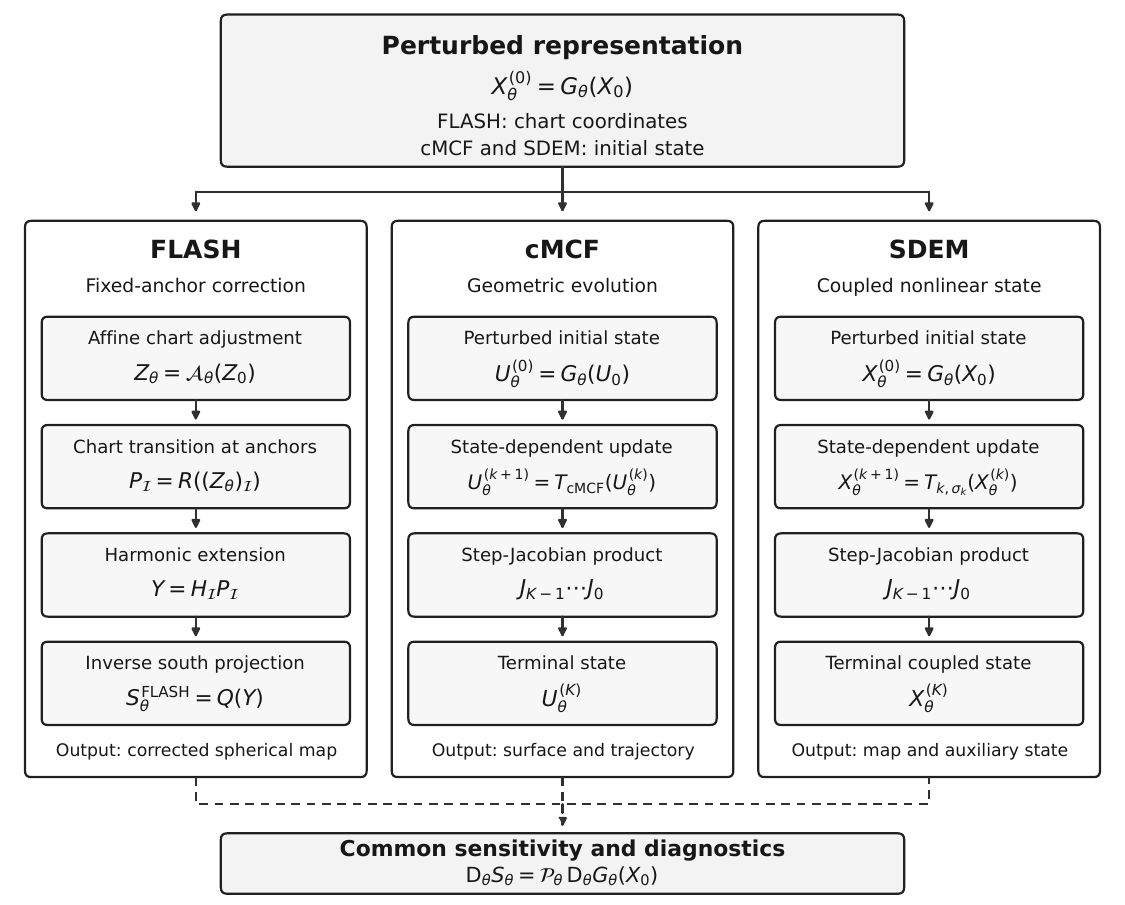}
\caption{\textbf{Unified view of representation-perturbation propagation in the three computational settings.} The perturbation acts on a chart or initial state, is propagated by the method-specific map or iteration, and is evaluated at the output through distortion, density, or downstream behavior.}
\label{fig:representation-framework}
\end{figure}

Figure~\ref{fig:representation-framework} summarizes the common structure used in the three numerical settings. A perturbation may act on the planar chart in FLASH~\cite{ChoiLamLui2015FLASH}, on the initial state of cMCF~\cite{KazhdanSolomonBenChen2012}, or on the coupled initial state of SDEM~\cite{LyuLuiChoi2024SDEM}. Each method then propagates that perturbation through its own active map, and the resulting output is evaluated through geometric distortion, density, or downstream behavior. The figure also makes clear that the perturbation location and the propagation operator are separate objects: the same output diagnostic can receive changes through different computational paths.

For a nonzero input perturbation $v$, define
\begin{equation}
 a_\theta(v)=\frac{\|\mathcal P_\theta v\|_2}{\|v\|_2},\qquad
 \mathfrak A(\theta)=\|\mathcal P_\theta\|_2
 =\max_{v\ne0}a_\theta(v).
 \label{eq:rs-amplification}
\end{equation}
For a differentiable scalar diagnostic $\mathcal J$, its response is
\begin{equation}
 \mathrm D_\theta(\mathcal J\circ S)_\theta
 =\mathrm D\mathcal J(S_\theta)\mathcal P_\theta\,
   \mathrm D_\theta G_\theta(X_0).
 \label{eq:rs-objective-chain}
\end{equation}
State amplification and distortion reduction are consequently different quantities. Section~\ref{subsec:rs-branch-stability} gives a sufficient neighborhood in which strict branch-selection inequalities remain satisfied. For FLASH, we also derive a specific radius-gap condition for anchor stability.

\subsection{A sufficient neighborhood for branch stability}
\label{subsec:rs-branch-stability}

The differential description above applies only while the discrete regime remains unchanged. To quantify this restriction, suppose that the branch at step $k$ is characterized by strict inequalities
\begin{equation}
    g_{k,r}\!\left(X^{(k)}\right)>0,
    \qquad r=1,\ldots,m_k.
    \label{eq:rs-branch-ineq}
\end{equation}
At the baseline parameter $\theta_0$, let  $\gamma_k
    :=\underset{1\le r\le m_k}{\min}
    g_{k,r}\!\left(X^{(k)}_{\theta_0}\right)>0$. Assume that, in a neighborhood of the baseline trajectory,
\begin{equation}
    \left|g_{k,r}(X)-g_{k,r}(Y)\right|
    \le L_k\|X-Y\|_2,
    \label{eq:rs-g-lipschitz}
\end{equation}
and that
\begin{equation}
    \left\|X^{(k)}_\theta-X^{(k)}_{\theta_0}\right\|_2
    \le C_k\|\theta-\theta_0\|_2.
    \label{eq:rs-state-lipschitz}
\end{equation}
We have the following result:
\begin{proposition}[Sufficient branch-stability radius]
\label{prop:rs-stability-radius}
Under \eqref{eq:rs-branch-ineq}--\eqref{eq:rs-state-lipschitz}, the complete branch sequence is unchanged within the neighborhood on which the bounds hold whenever
\begin{equation}
    \|\theta-\theta_0\|_2<r_{\mathrm{stab}},
    \qquad
    r_{\mathrm{stab}}
    :=\min_{0\le k<K}\frac{\gamma_k}{L_kC_k}.
    \label{eq:rs-stability-radius}
\end{equation}
A ratio with $L_kC_k=0$ is interpreted as $+\infty$. If $G_\theta$ is locally Lipschitz in $\theta$ with constant $L_G$ and the active update at step $j$ is locally Lipschitz with constant $L_{T_j}$, one may take
\begin{equation}
    C_k\le L_G\prod_{j=0}^{k-1}L_{T_j},
    \qquad k\ge 1.
    \label{eq:rs-Ck-product}
\end{equation}
\end{proposition}

\begin{proof}
For every active inequality,
\[
    g_{k,r}\!\left(X^{(k)}_\theta\right)
    \ge
    g_{k,r}\!\left(X^{(k)}_{\theta_0}\right)
    -L_k\left\|X^{(k)}_\theta-X^{(k)}_{\theta_0}\right\|_2
    \ge
    \gamma_k-L_kC_k\|\theta-\theta_0\|_2.
\]
The right-hand side is positive under \eqref{eq:rs-stability-radius}, so no switching surface is crossed. The product estimate \eqref{eq:rs-Ck-product} follows by repeated application of the local Lipschitz bounds along the baseline branch.
\end{proof}

The radius in \eqref{eq:rs-stability-radius} is a sufficient bound. It distinguishes smooth propagation within a fixed branch from changes caused by crossing a switching surface. The FLASH-specific radius-gap condition is given in Proposition~\ref{prop:flash-anchor-stability}.

\section{Exact chart dependence of the FLASH correction}
\label{sec:flash-exact}

FLASH~\cite{ChoiLamLui2015FLASH} first constructs a planar north-chart realization and applies its north--south size balancing procedure. Let $Z_0\in\mathbb{R}^{n\times 2}$ denote the resulting coordinates, which serve as the baseline representation. All subsequent FLASH operations are held fixed as this representation is perturbed.

For the optimization study, we use the five-parameter affine family
\begin{equation}
    Z_\theta=\mathcal A_\theta(Z_0),
    \qquad
    \mathcal A_{a,B,t}(z)=\mathrm e^a\mathrm e^Bz+t,
    \label{eq:flash-affine-family}
\end{equation}
where
\begin{equation}
    B=
    \begin{pmatrix}
        b_1 & b_2\\
        b_2 & -b_1
    \end{pmatrix}
    \in\operatorname{Sym}_0(2),
    \qquad
    \theta=(a,b_1,b_2,t_1,t_2).
    \label{eq:flash-B}
\end{equation}
The parameters control isotropic scale, anisotropy, and translation. Section~\ref{sec:preconditioning} specifies their bounds and the search procedure. The following analysis first treats the geometric lift and then the correction.

\subsection{Finite stereographic geometry}
\label{subsec:flash-stereo-geometry}

We first examine representation sensitivity before the south-pole correction.  For $z=(x,y)\in\mathbb{R}^2$, the inverse north stereographic projection is
\begin{equation}
    \Pi_N^{-1}(z)
    =\frac{1}{1+\|z\|_2^2}
    \begin{pmatrix}
        2x\\ 2y\\ \|z\|_2^2-1
    \end{pmatrix}.
    \label{eq:flash-inv-stereo}
\end{equation}
For a planar vertex $z_i$, set $q_i=1+\|z_i\|_2^2$ and $m_{ij}=\frac{2}{\sqrt{q_iq_j}}$. We have the following result:
\begin{proposition}[Exact finite-edge scaling]
\label{prop:flash-finite-edge}
For any two finite planar points $z_i,z_j$,
\begin{equation}
    \left\|\Pi_N^{-1}(z_i)-\Pi_N^{-1}(z_j)\right\|_2
    =m_{ij}\|z_i-z_j\|_2.
    \label{eq:flash-exact-edge}
\end{equation}
For a face $T=(i,j,k)$, define its logarithmic radial spread by
\begin{equation}
    \chi_T(Z)
    :=\max_{r\in T}\log(1+\|z_r\|_2^2)
     -\min_{r\in T}\log(1+\|z_r\|_2^2).
    \label{eq:flash-radial-spread}
\end{equation}
Then the three edge multipliers satisfy
\begin{equation}
    \max_{e,e'\subset T}
    \left|\log\frac{m_e}{m_{e'}}\right|
    =\frac12\chi_T(Z).
    \label{eq:flash-edge-scale-range}
\end{equation}
In particular, equal vertex radii make the spherical chord triangle exactly similar to the planar triangle.
\end{proposition}

\begin{proof}
Writing $s_r=\Pi_N^{-1}(z_r)$, direct substitution gives
\begin{equation}
    \|s_i-s_j\|_2^2
    =2(1-s_i^Ts_j)
    =\frac{4\|z_i-z_j\|_2^2}{q_iq_j},
\end{equation}
which proves \eqref{eq:flash-exact-edge}.  If $a_r=\log q_r$ and $a_i\le a_j\le a_k$, the largest and smallest logarithmic edge multipliers are $\log 2-(a_i+a_j)/2$ and $\log 2-(a_j+a_k)/2$.  Their difference is
$(a_k-a_i)/2$.
\end{proof}

To convert this exact edge identity into a shape estimate, fix a face $T=(i,j,k)$.  Let $B_T\in\mathbb{R}^{2\times2}$ be any nonsingular local edge matrix for the source triangle and let
\begin{equation}
    D_T(Z)=[z_j-z_i,\ z_k-z_i],
    \qquad
    G_T(Z)=D_T(Z)^TD_T(Z).
    \label{eq:flash-planar-gram}
\end{equation}
For a spherical realization $S$, let $C_T(S)\in\mathbb{R}^{2\times2}$ be any planar realization of its chord triangle with the same three chord lengths.  The local log-condition distortion is
\begin{equation}
    d_T(S)=\log\kappa_2\!\left(C_T(S)B_T^{-1}\right),
    \qquad
    d_T(Z)=\log\kappa_2\!\left(D_T(Z)B_T^{-1}\right).
    \label{eq:flash-local-logcond}
\end{equation}
This quantity is invariant under rigid changes of the local coordinates used for $C_T(S)$.

\begin{theorem}[Radial variation and finite-triangle conditioning]
\label{thm:flash-radial-conditioning}
Let
\begin{equation}
    h_T=\max_{r,s\in T}\|z_r-z_s\|_2,
    \qquad
    \delta_T=\exp\!\left(\frac{\chi_T(Z)}{2}\right)-1,
    \label{eq:flash-delta}
\end{equation}
and define
\begin{equation}
    \varepsilon_T
    =\frac{5h_T^2}{2\lambda_{\min}(G_T(Z))}\,\delta_T.
    \label{eq:flash-epsilon}
\end{equation}
If $\varepsilon_T<1$, then
\begin{equation}
    \left|d_T\!\left(\Pi_N^{-1}(Z)\right)-d_T(Z)\right|
    \le
    \frac12\log\frac{1+\varepsilon_T}{1-\varepsilon_T}.
    \label{eq:flash-conditioning-bound}
\end{equation}
\end{theorem}

\begin{proof}
Choose $m_*>0$ so that $\log m_*$ is the midpoint of the extreme logarithmic edge multipliers on $T$.  By Proposition~\ref{prop:flash-finite-edge}, $|\log(m_e/m_*)|\le\chi_T/4$.  After dividing all target chord lengths by $m_*$, the squared length of every edge changes from its planar value by at most $\delta_T h_T^2$.

The diagonal entries of an edge Gram matrix are squared side lengths, and the off-diagonal entry is one half of the sum of two squared side lengths minus the third.  Hence the normalized target Gram matrix $\widetilde G_T$ obeys
\begin{equation}
    \|\widetilde G_T-G_T\|_2
    \le \frac52\delta_Th_T^2,
\end{equation}
and therefore
\begin{equation}
    (1-\varepsilon_T)G_T
    \preceq \widetilde G_T
    \preceq (1+\varepsilon_T)G_T.
\end{equation}
If $K_T$ denotes the affine transfer from the planar triangle to the normalized target chord triangle, congruence by $D_T^{-1}$ gives
\begin{equation}
    1-\varepsilon_T
    \le\sigma_{\min}(K_T)^2
    \le\sigma_{\max}(K_T)^2
    \le1+\varepsilon_T.
\end{equation}
The source-to-target transfer is, up to a rigid target coordinate choice and a common scale, the product of $K_T$ and the source-to-planar transfer.  Condition-number submultiplicativity applied to this factorization and its inverse yields
\eqref{eq:flash-conditioning-bound}.
\end{proof}

The bound depends on both planar element shape, through $\lambda_{\min}(G_T)$, and radial variation, through $\chi_T$. Thus, smooth conformality does not eliminate the dependence of finite-triangle conditioning on the planar representation. We next examine how the south-pole correction changes this dependence.

\subsection{Exact reduction of the south-pole Beltrami correction}
\label{subsec:flash-lbs-reduction}

The calculation below specializes the metric pullback of the Dirichlet form to the LBS tensor used in FLASH~\cite{ChoiLamLui2015FLASH}; the surface finite-element background can be found in \cite{DziukElliott2013}.

Let $P=(p_1,\ldots,p_n)^T\in\mathbb{R}^{n\times2}$ be any finite planar realization with the same triangular connectivity $F$ as the source mesh $V$, and assume each planar face is nondegenerate.  On a face $T$, let $X_T:P_T\rightarrow V_T$ denote the affine map from the planar triangle to the corresponding source triangle.  In planar coordinates, its Jacobian is $J_T\in\mathbb{R}^{3\times2}$.
Define
\begin{equation}
    \mathsf G_T=J_T^TJ_T
    =\begin{pmatrix}E_T&F_T\\F_T&G_T^{(2)}\end{pmatrix},
    \qquad
    \Delta_T=\sqrt{E_TG_T^{(2)}-F_T^2}>0.
    \label{eq:flash-pullback-metric}
\end{equation}
The facewise Beltrami coefficient determined by this metric is
\begin{equation}
    \mu_T
    =\frac{E_T-G_T^{(2)}+2\mathrm i F_T}
    {E_T+G_T^{(2)}+2\Delta_T}.
    \label{eq:flash-metric-mu}
\end{equation}
For $\mu=\rho+\mathrm i\tau$, $|\mu|<1$, the coefficient tensor of the linear Beltrami solver (LBS)~\cite{LuiLamWongGu2013LBS} is
\begin{equation}
    \mathcal A(\mu)
    =\frac{1}{1-|\mu|^2}
    \begin{pmatrix}
        1-2\rho+|\mu|^2 & -2\tau\\
        -2\tau & 1+2\rho+|\mu|^2
    \end{pmatrix}.
    \label{eq:flash-lbs-tensor}
\end{equation}

\begin{lemma}[Metric form of the LBS tensor]
\label{lem:flash-lbs-metric}
For \eqref{eq:flash-metric-mu},
\begin{equation}
    \mathcal A(\mu_T)=\Delta_T\mathsf G_T^{-1}.
    \label{eq:flash-lbs-metric-id}
\end{equation}
\end{lemma}

\begin{proof}
Set $\tau_T=E_T+G_T^{(2)}+2\Delta_T$.  Direct calculation gives
\begin{equation}
    1-|\mu_T|^2=\frac{4\Delta_T}{\tau_T},
\end{equation}
together with
\begin{equation}
    \frac{|1-\mu_T|^2}{1-|\mu_T|^2}=\frac{G_T^{(2)}}{\Delta_T},
    \qquad
    \frac{-2\operatorname{Im}\mu_T}{1-|\mu_T|^2}=-\frac{F_T}{\Delta_T},
    \qquad
    \frac{|1+\mu_T|^2}{1-|\mu_T|^2}=\frac{E_T}{\Delta_T}.
\end{equation}
Substitution into \eqref{eq:flash-lbs-tensor} yields
\begin{equation}
    \mathcal A(\mu_T)
    =\frac{1}{\Delta_T}
      \begin{pmatrix}G_T^{(2)}&-F_T\\-F_T&E_T\end{pmatrix}
    =\Delta_T\mathsf G_T^{-1}.
\end{equation}
\end{proof}

For a scalar piecewise-linear function $u$ on $P_T$, let $\widehat u=u\circ X_T^{-1}$ denote the corresponding function on $V_T$. We have the following result:

\begin{theorem}[Exact harmonic reduction of the unmodified LBS correction]
\label{thm:flash-lbs-reduction}
Let the facewise Beltrami coefficients be computed from the affine maps $P_T\to V_T$ by \eqref{eq:flash-metric-mu}.  Assume that the same connectivity, piecewise-linear basis, and facewise integration convention are used in the LBS assembly and in the source Dirichlet form.  Then
\begin{equation}
    \sum_{T\in F}\int_{P_T}
    \nabla u^T\mathcal A(\mu_T)\nabla v\,\mathrm dp
    =
    \sum_{T\in F}\int_{V_T}
    \nabla_V\widehat u\cdot\nabla_V\widehat v\,\mathrm dA.
    \label{eq:flash-energy-equivalence}
\end{equation}
Consequently, under the standard finite-element normalization,
\begin{equation}
    K_{\mathrm{LBS}}=K_V,
    \label{eq:flash-stiffness-equivalence}
\end{equation}
where $K_V$ is the intrinsic piecewise-linear Dirichlet stiffness matrix of the source mesh \cite{PinkallPolthier1993,MeyerDesbrunSchroderBarr2003}. A uniform nonzero global scalar or sign convention does not change the associated homogeneous Dirichlet solution.
\end{theorem}

\begin{proof}
The area factor induced by $X_T$ is $\mathrm dA=\Delta_T\,\mathrm dp$, while the pullback of the intrinsic gradient satisfies
\begin{equation}
    \nabla_V\widehat u\cdot\nabla_V\widehat v
    =\nabla u^T\mathsf G_T^{-1}\nabla v.
\end{equation}
Lemma~\ref{lem:flash-lbs-metric} therefore gives equality of the two facewise bilinear forms. Summing over $T\in F$ proves \eqref{eq:flash-energy-equivalence}.  Equality of the assembled matrices follows by evaluating the bilinear forms in the nodal basis.
\end{proof}

Theorem~\ref{thm:flash-lbs-reduction} applies to the unmodified assembly described above. Clipping or smoothing the coefficients, changing the weights or connectivity, or using a different integration rule may alter the operator and invalidate
\eqref{eq:flash-stiffness-equivalence}.

Let $\mathcal I$ be the set of vertices prescribed by the south-pole correction and let $\mathcal U$ be the complement.  Partitioning $K_V$ accordingly and we have the following:

\begin{corollary}[Source-mesh harmonic extension]
\label{cor:flash-harmonic-extension}
Assume $\mathcal I\neq\varnothing$.  If $P_{\mathcal I}$ denotes the prescribed south-plane coordinates, then the exact correction satisfies
\begin{equation}
    Y_{\mathcal I}=P_{\mathcal I},
    \qquad
    Y_{\mathcal U}
    =-K_{\mathcal U\mathcal U}^{-1}
      K_{\mathcal U\mathcal I}P_{\mathcal I}.
    \label{eq:flash-harmonic-block}
\end{equation}
Equivalently, $Y=H_{\mathcal I}P_{\mathcal I}$,
where $H_{\mathcal I}$ depends only on the source mesh and the prescribed index set.
\end{corollary}

\begin{proof}
For a connected mesh, the Dirichlet block $K_{\mathcal U\mathcal U}$ is positive definite when the prescribed set is nonempty.  The free Euler--Lagrange equations for the Dirichlet energy therefore give \eqref{eq:flash-harmonic-block}; stacking the prescribed and free rows defines $H_{\mathcal I}$.
\end{proof}

Although the intermediate Beltrami coefficients depend on every planar face, this dependence cancels in the assembled stiffness matrix. For a fixed anchor set, the corrected planar map therefore depends on the incoming representation only through the prescribed south-plane coordinates.

\subsection{Exact branchwise form of the corrected FLASH map}
\label{subsec:flash-closed-form}

Let $P_S$ denote south stereographic projection and $Q=P_S^{-1}$ its inverse.  With the conventions in \eqref{eq:flash-inv-stereo},
\begin{equation}
    R(z):=(P_S\circ\Pi_N^{-1})(z)=\frac{z}{\|z\|_2^2},
    \qquad z\neq0,
    \label{eq:flash-chart-transition}
\end{equation}
and
\begin{equation}
    Q(y)
    =\frac{1}{1+\|y\|_2^2}
      \begin{pmatrix}2y_1\\2y_2\\1-\|y\|_2^2\end{pmatrix}.
    \label{eq:flash-Q}
\end{equation}
The height of the raw spherical point $\Pi_N^{-1}(z_i)$ is
\begin{equation}
    h_i(Z)=\frac{\|z_i\|_2^2-1}{1+\|z_i\|_2^2},
    \label{eq:flash-height-radius}
\end{equation}
which is strictly increasing in $\|z_i\|_2^2$.  Hence selecting the $k$ lowest raw spherical vertices is equivalent to selecting the $k$ smallest planar squared radii.

Fix $\theta_0$, and let $\mathcal I$ be the anchor set selected at that parameter value.  Define the radius gap
\begin{equation}
    \gamma_{\mathrm{FLASH}}
    =\min_{j\notin\mathcal I}\|Z_{\theta_0,j}\|_2^2
     -\max_{i\in\mathcal I}\|Z_{\theta_0,i}\|_2^2.
    \label{eq:flash-radius-gap}
\end{equation}

\begin{proposition}[FLASH anchor-set stability]
\label{prop:flash-anchor-stability}
Suppose $\gamma_{\mathrm{FLASH}}>0$ and, in a neighborhood of $\theta_0$,
\begin{equation}
    \left|
    \|Z_{\theta,i}\|_2^2-\|Z_{\theta_0,i}\|_2^2
    \right|
    \le L_{\mathrm{rad}}\|\theta-\theta_0\|_2
    \qquad\text{for every }i.
    \label{eq:flash-radius-Lipschitz}
\end{equation}
Then
\begin{equation}
    \|\theta-\theta_0\|_2
    <\frac{\gamma_{\mathrm{FLASH}}}{2L_{\mathrm{rad}}}
    \quad\Longrightarrow\quad
    \mathcal I(\theta)=\mathcal I(\theta_0).
    \label{eq:flash-anchor-radius}
\end{equation}
\end{proposition}

\begin{proof}
For $i\in\mathcal I$ and $j\notin\mathcal I$,
\begin{equation}
    \|Z_{\theta,j}\|_2^2-\|Z_{\theta,i}\|_2^2
    \ge
    \gamma_{\mathrm{FLASH}}
    -2L_{\mathrm{rad}}\|\theta-\theta_0\|_2>0.
\end{equation}
Thus no selected and unselected radius can exchange order.
\end{proof}

\begin{theorem}[Exact fixed-anchor representation of FLASH]
\label{thm:flash-closed-form}
Assume the hypotheses of Proposition~\ref{prop:flash-anchor-stability}; assume also that every selected $Z_{\theta_0,i}$ is nonzero and that the south-plane faces used by the correction are nondegenerate. On a sufficiently small fixed-anchor neighborhood in which the unmodified LBS branch is used,
\begin{equation}
    S^{\mathrm{FLASH}}_\theta
    =Q\!\left(
      H_{\mathcal I}
      R\!\left((Z_\theta)_{\mathcal I}\right)
    \right),
    \label{eq:flash-closed-form}
\end{equation}
with $R$ and $Q$ applied rowwise.  In particular, $S^{\mathrm{FLASH}}_\theta$ is real analytic in the physical affine parameters on that branch.
\end{theorem}

\begin{proof}
Equation \eqref{eq:flash-height-radius} and Proposition~\ref{prop:flash-anchor-stability} keep the selected set equal to $\mathcal I$.  The prescribed south-plane data are therefore $P_{\mathcal I}=R((Z_\theta)_{\mathcal I})$.  Corollary~\ref{cor:flash-harmonic-extension} gives $Y=H_{\mathcal I}P_{\mathcal I}$, and applying \eqref{eq:flash-Q} gives \eqref{eq:flash-closed-form}.  The affine matrix exponential, $R$ away from the origin, the linear harmonic map, and $Q$ are analytic.
\end{proof}

The corrected FLASH map therefore has the following exact representation on a fixed branch:
\begin{equation}
    \mathcal A_\theta
    \;\longrightarrow\;
    R
    \;\longrightarrow\;
    H_{\mathcal I}
    \;\longrightarrow\;
    Q.
    \label{eq:flash-propagation-chain}
\end{equation}
This representation applies while the anchors and the unmodified correction branch remain fixed and the chart configurations remain nondegenerate. Anchor switching, coefficient modification, degeneracy, and fallback events require separate treatment.

\subsection{Pole-aware propagation operator}
\label{subsec:flash-propagator}

For $z\neq0$,
\begin{equation}
    \mathrm{D} R(z)
    =\frac{1}{\|z\|_2^2}
      \left(I_2-2\widehat z\widehat z^T\right),
    \qquad
    \widehat z=\frac{z}{\|z\|_2},
    \qquad
    \|\mathrm{D} R(z)\|_2=\|z\|_2^{-2}.
    \label{eq:flash-DR}
\end{equation}
For \eqref{eq:flash-Q},
\begin{equation}
    \mathrm{D} Q(y)^T\mathrm{D} Q(y)
    =\frac{4}{(1+\|y\|_2^2)^2}I_2,
    \qquad
    \|\mathrm{D} Q(y)\|_2=\frac{2}{1+\|y\|_2^2}.
    \label{eq:flash-DQ}
\end{equation}

Let $P_{\mathcal I}=R((Z_\theta)_{\mathcal I})$, $Y=H_{\mathcal I}P_{\mathcal I}$, and stack vertex coordinates row by row when applying derivatives. Define the block-diagonal derivatives
\begin{equation}
    \mathcal R_{Z_{\mathcal I}}
    =\operatorname{blkdiag}\{\mathrm{D} R(Z_i):i\in\mathcal I\},
    \qquad
    \mathcal Q_Y
    =\operatorname{blkdiag}\{\mathrm{D} Q(Y_j):1\le j\le n\}.
    \label{eq:flash-block-derivatives}
\end{equation}
We have the following theorem:
\begin{theorem}[Jacobian factorization of the corrected map]
\label{thm:flash-jacobian}
Within the fixed-anchor regime of Theorem~\ref{thm:flash-closed-form}, any parameter direction $\xi$ satisfies
\begin{equation}
    \mathrm{D} S^{\mathrm{FLASH}}_\theta[\xi]
    =
    \mathcal Q_Y
    (H_{\mathcal I}\otimes I_2)
    \mathcal R_{Z_{\mathcal I}}
    \mathrm{D} Z_{\theta,\mathcal I}[\xi].
    \label{eq:flash-jacobian}
\end{equation}
Hence the solver-side FLASH propagator is
\begin{equation}
    \mathcal P_{\mathrm{FLASH}}
    =\mathcal Q_Y
      (H_{\mathcal I}\otimes I_2)
      \mathcal R_{Z_{\mathcal I}},
    \label{eq:flash-propagator}
\end{equation}
and obeys
\begin{equation}
    \|\mathcal P_{\mathrm{FLASH}}\|_2
    \le
    2\|H_{\mathcal I}\otimes I_2\|_2
    \max_{i\in\mathcal I}\|Z_i\|_2^{-2}.
    \label{eq:flash-propagator-bound}
\end{equation}
\end{theorem}

\begin{proof}
Differentiate \eqref{eq:flash-closed-form} and note that the matrix $H_{\mathcal I}$ is constant on the fixed-anchor branch, giving \eqref{eq:flash-jacobian}.  The norm estimate follows from \eqref{eq:flash-DR}, \eqref{eq:flash-DQ}, and submultiplicativity.
\end{proof}

The factorization separates three contributions to pole sensitivity. Small anchor radii can produce large chart-transition derivatives in $\mathcal R_{Z_{\mathcal I}}$. The harmonic extension propagates these perturbations over the mesh, and $\mathcal Q_Y$ applies the local inverse-stereographic scale. Equation~\eqref{eq:flash-propagator-bound} is an upper bound: cancellation between the factors may yield a much smaller directional response than the product of their norms suggests.

For the affine family \eqref{eq:flash-affine-family}, let
$\xi=(\dot a,\dot B,\dot t)$.  The representation derivative at a vertex is
\begin{equation}
    \mathrm{D} Z_i[\xi]
    =\mathrm e^a
      \left(\dot a\,\mathrm e^B
      +\mathrm{D}(\exp)_B[\dot B]\right)z_i+\dot t,
    \label{eq:flash-affine-derivative}
\end{equation}
where the Fr\'echet derivative of the matrix exponential~\cite{Higham2008Functions} is
\begin{equation}
    \mathrm{D}(\exp)_B[\dot B]
    =\int_0^1
      \mathrm e^{(1-s)B}\dot B\,\mathrm e^{sB}\,\mathrm ds.
    \label{eq:flash-dexp}
\end{equation}
At the identity preconditioner,
\begin{equation}
    \mathrm{D} Z_i[\xi]
    =(\dot a I_2+\dot B)z_i+\dot t.
    \label{eq:flash-affine-derivative-zero}
\end{equation}
At the identity, translation produces a planar displacement independent of the vertex radius, whereas scale and anisotropy produce displacements proportional to that radius. The chart transition then acts on these displacements through its $\|Z_i\|^{-2}$ derivative.

\begin{corollary}[Cancellation at prescribed vertices]
\label{cor:flash-anchor-cancellation}
For every $i\in\mathcal I$,
\begin{equation}
    S_i^{\mathrm{FLASH}}
    =Q(R(Z_i))=\Pi_N^{-1}(Z_i),
    \qquad
    \|\mathrm{D}(Q\circ R)(Z_i)\|_2
    =\frac{2}{1+\|Z_i\|_2^2}.
    \label{eq:flash-anchor-cancellation}
\end{equation}
For any output vertex $j$ and parameter direction $\xi$,
\begin{equation}
    \left\|\mathrm{D} S_j^{\mathrm{FLASH}}[\xi]\right\|_2
    \le
    \frac{2}{1+\|Y_j\|_2^2}
    \sum_{i\in\mathcal I}
    |(H_{\mathcal I})_{ji}|\,
    \|Z_i\|_2^{-2}\,
    \|\mathrm{D} Z_i[\xi]\|_2.
    \label{eq:flash-local-bound}
\end{equation}
\end{corollary}

\begin{proof}
The prescribed rows of $H_{\mathcal I}$ are identity rows, which gives the first identity. The derivative norm follows by composing \eqref{eq:flash-DR} and \eqref{eq:flash-DQ}.  Applying the triangle inequality to the $j$th row of \eqref{eq:flash-jacobian} gives
\eqref{eq:flash-local-bound}.
\end{proof}

The intermediate $\|Z_i\|^{-2}$ amplification cancels exactly at prescribed spherical vertices, but not generally at free vertices.  The free-vertex response depends on the harmonic weights and on the inverse-projection scale.  On a general cotangent mesh these harmonic weights need not be nonnegative, so \eqref{eq:flash-local-bound} is stated with absolute values rather than as a convex-combination estimate \cite{WardetzkyMathurKaelbererGrinspun2007}.

\subsection{Propagation to element distortion}
\label{subsec:flash-distortion-sensitivity}

The state-space response in \eqref{eq:flash-jacobian} also determines the first-order change in element shape.  For a face $T=(i,j,k)$ define
\begin{equation}
    E_T(S)=[s_j-s_i,\ s_k-s_i]\in\mathbb{R}^{3\times2},
    \qquad
    M_T(S)=B_T^{-T}E_T(S)^TE_T(S)B_T^{-1}.
    \label{eq:flash-MT}
\end{equation}
The eigenvalues of $M_T$ are the squared singular values of the affine transfer from the source triangle to the target chord triangle.  Thus, if
$\lambda_1\ge\lambda_2>0$,
\begin{equation}
    d_T(S)=\frac12\log\frac{\lambda_1(M_T(S))}{\lambda_2(M_T(S))}.
    \label{eq:flash-d-eigen}
\end{equation}
For a perturbation $\dot S$, let
\begin{equation}
    \dot E_T=[\dot s_j-\dot s_i,\ \dot s_k-\dot s_i],
    \qquad
    \dot M_T
    =B_T^{-T}
      (\dot E_T^TE_T+E_T^T\dot E_T)
      B_T^{-1}.
    \label{eq:flash-dotM}
\end{equation}
We have the following result:
\begin{proposition}[Directional sensitivity of corrected element conditioning]
\label{prop:flash-distortion-derivative}
Suppose $M_T(S)$ is positive definite.  If $\lambda_1>\lambda_2$ with normalized eigenvectors $q_1,q_2$, then
\begin{equation}
    \mathrm{D} d_T(S)[\dot S]
    =\frac12\left(
       \frac{q_1^T\dot M_Tq_1}{\lambda_1}
       -\frac{q_2^T\dot M_Tq_2}{\lambda_2}
    \right).
    \label{eq:flash-d-derivative}
\end{equation}
If $M_T(S)=\lambda I_2$, the one-sided directional derivative is
\begin{equation}
    d_T'(S;\dot S)
    =\frac{1}{2\lambda}
      \left[\lambda_{\max}(\dot M_T)-\lambda_{\min}(\dot M_T)\right].
    \label{eq:flash-d-isotropic}
\end{equation}
Combining either formula with \eqref{eq:flash-jacobian} gives the first-order response of the fully corrected element distortion to the planar representation parameters.
\end{proposition}

\begin{proof}
Equation~\eqref{eq:flash-d-eigen} follows from $M_T=J_T^TJ_T$ for the source-to-chord transfer. For distinct eigenvalues, standard spectral differentiation gives $\dot\lambda_r=q_r^T\dot M_Tq_r$~\cite{Lewis1996Spectral}; differentiating~\eqref{eq:flash-d-eigen} then yields~\eqref{eq:flash-d-derivative}. At $M_T=\lambda I_2$, the one-sided directional derivatives of the ordered eigenvalues are the ordered eigenvalues of $\dot M_T$~\cite{Lewis1999Eigenvalues}, yielding~\eqref{eq:flash-d-isotropic}.
\end{proof}

\section{Representation sensitivity in cMCF and SDEM}
\label{sec:other-realizations}

We apply the framework of Section~\ref{sec:rs-framework} to two iterative methods. cMCF exposes sensitivity through geometric updates, whereas SDEM includes feedback through auxiliary state variables. The formulas below describe how an initial perturbation propagates through each iteration; the FLASH operator reduction is specific to the preceding section.

\subsection{cMCF: smooth state-dependent geometric propagation}
\label{subsec:cmcf-realization}

In conformalized mean-curvature flow (cMCF) \cite{KazhdanSolomonBenChen2012}, the linear system at each step depends on the current geometry. For a fixed iteration count, the update is a smooth state-dependent map. Let $U^{(k)}\in\mathbb{R}^{n\times 3}$ denote the evolving embedding. The baseline discretization holds the initial stiffness matrix $L_0$ fixed and recomputes the mass matrix from the
current state,
\begin{equation}
    M_k=M\!\left(U^{(k)}\right).
    \label{eq:cmcf-mass}
\end{equation}
For a fixed time step $\delta>0$, write the unnormalized implicit update as
\begin{equation}
    \widehat U^{(k+1)}
    =\Phi\!\left(U^{(k)}\right)
    :=\left(M_k-\delta L_0\right)^{-1}M_kU^{(k)}.
    \label{eq:cmcf-implicit-step}
\end{equation}
Let $\mathcal N$ denote the deterministic recentering and normalization applied after the implicit solve. One cMCF step is therefore
\begin{equation}
    U^{(k+1)}
    =T_{\mathrm{cMCF}}\!\left(U^{(k)}\right)
    :=\mathcal N\!\left(\Phi\!\left(U^{(k)}\right)\right).
    \label{eq:cmcf-update}
\end{equation}
The representation perturbation is introduced only through the initial state,
\begin{equation}
    U^{(0)}_\theta=G_\theta(U_0),
    \label{eq:cmcf-initial-family}
\end{equation}
while $L_0$, $\delta$, the normalization rule, and all other solver settings are held fixed. This protocol isolates sensitivity to the initial representation.

The local response of one implicit step can be written explicitly.  Define $A(U):=M(U)-\delta L_0$ and $\widehat U=\Phi(U)$. For a tangent direction $V$, differentiation of $A(U)\widehat U=M(U)U$ gives
\begin{equation}
    \mathrm{D}\Phi(U)[V]
    =A(U)^{-1}
      \left[
        M(U)V+\mathrm{D} M(U)[V]\,(U-\widehat U)
      \right].
    \label{eq:cmcf-step-derivative}
\end{equation}
Hence the normalized step Jacobian is
\begin{equation}
    J_k^{\mathrm{cMCF}}
    =\mathrm{D}\mathcal N\!\left(\widehat U^{(k+1)}\right)
      \mathrm{D}\Phi\!\left(U^{(k)}\right),
    \label{eq:cmcf-Jk}
\end{equation}
and a tangent perturbation obeys
\begin{equation}
    \dot U^{(k+1)}
    =J_k^{\mathrm{cMCF}}\dot U^{(k)}.
    \label{eq:cmcf-tangent-recursion}
\end{equation}
For a fixed iteration count $K$,
\begin{equation}
    \mathrm{D}_\theta U^{(K)}_\theta
    =J_{K-1}^{\mathrm{cMCF}}\cdots J_0^{\mathrm{cMCF}}
      \mathrm{D}_\theta G_\theta(U_0).
    \label{eq:cmcf-product}
\end{equation}
In cMCF, a perturbation of the embedding changes the mass matrix, which affects the next implicit solve and all subsequent normalized states. Its local response is therefore described by a product of step Jacobians, each accounting for the mass matrix, implicit solve, and normalization.

Equation~\eqref{eq:cmcf-product} describes the response at a fixed iteration count. For convergence-based stopping, the stopping index is an additional discrete variable. The numerical study reports this index separately and also compares trajectories at the baseline stopping horizon to examine propagation over a common number of steps.

\subsection{SDEM: nonlinear state-dependent propagation}
\label{subsec:sdem-realization}

For open surfaces, density-equalizing parameterizations have been developed using diffusion-based deformation~\cite{ChoiRycroft2018} and extended to incorporate quasiconformal distortion control~\cite{LyuChoiLui2024DEQ}. In the spherical SDEM method~\cite{LyuLuiChoi2024SDEM}, the evolving numerical state contains the spherical map together with any auxiliary variables needed by the nonlinear update. Denote this full state by $X^{(k)}$ and write $X^{(0)}_\theta=G_\theta(X_0)$, $ X^{(k+1)}_\theta
    =T_{k,\sigma_k}\!\left(X^{(k)}_\theta\right)$. The controlled perturbation is applied only to the initial representation; population size, time step, tolerance, maximum iteration count, and the remaining solver configuration are held fixed in the mechanism experiment.

Within a fixed computational regime, set     $J_k
    =\mathrm{D}_XT_{k,\sigma_k}\!\left(X^{(k)}_\theta\right)$. Then
\begin{equation}
    \mathrm{D}_\theta X^{(K)}_\theta
    =J_{K-1}\cdots J_0\,\mathrm{D}_\theta G_\theta(X_0),
    \label{eq:sdem-product}
\end{equation}
and the tangent response in a parameter direction $\xi$ satisfies the recursion
\begin{equation}
    \dot X^{(0)}=\mathrm{D}_\theta G_\theta(X_0)[\xi],
    \qquad
    \dot X^{(k+1)}=J_k\dot X^{(k)}.
    \label{eq:sdem-tangent-recursion}
\end{equation}
If the reported output is only the spherical component of the full state, the corresponding coordinate projection is appended to the left of \eqref{eq:sdem-product}.  Including the auxiliary variables captures sensitivity that passes through the coupled state before affecting the spherical map.

\section{Pole-aware planar preconditioning}
\label{sec:preconditioning}

The affine family in \eqref{eq:flash-affine-family} provides a small set of chart adjustments. Its determinant is positive, and its anisotropy can be bounded explicitly. We first quantify these parameter effects, then define the distortion measures and the candidate search. The final acceptance test uses the complete FLASH output; the raw stereographic objective is only a candidate-generation surrogate.

\begin{figure}[t]
\centering
\includegraphics[width=\linewidth]
{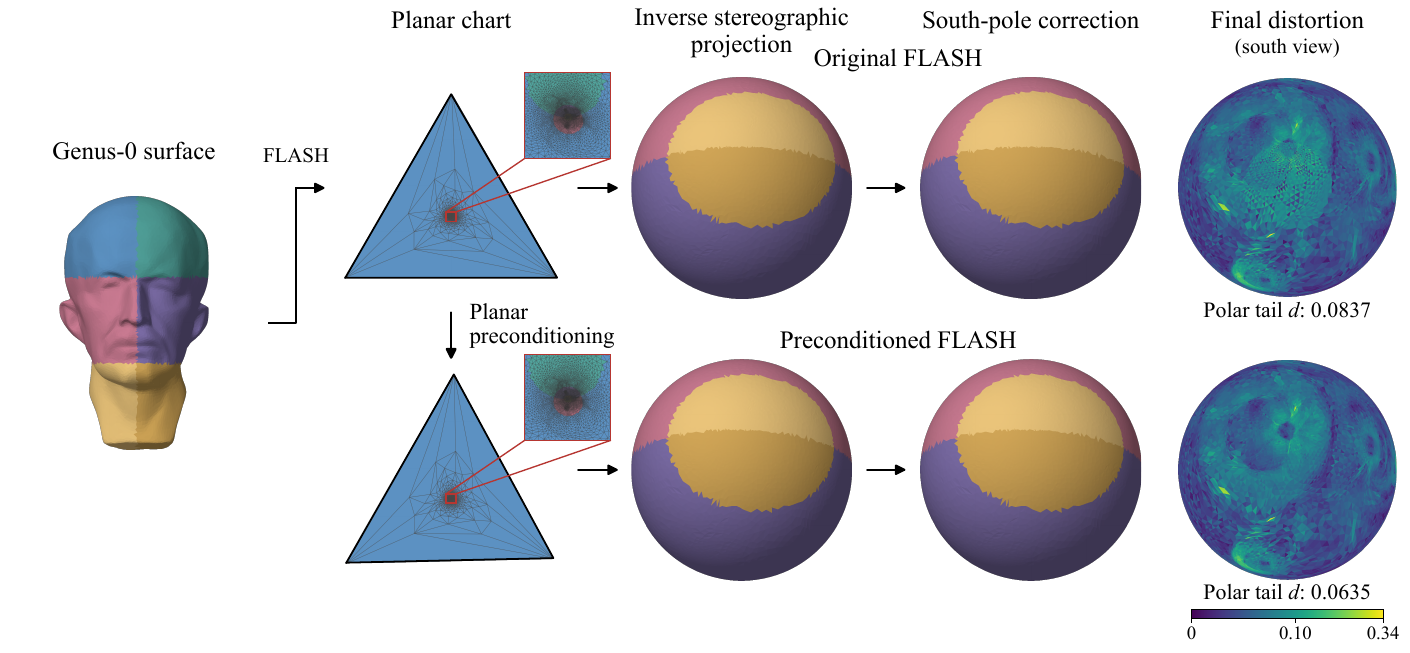}

\caption{
\textbf{The proposed planar preconditioning applied to the FLASH method~\cite{ChoiLamLui2015FLASH} for the spherical conformal parameterization of the Max Planck surface.}
The lower branch inserts planar preconditioning before inverse stereographic projection; the upper branch shows the original FLASH. Both share the source mesh and south-pole correction rule. Upper-right insets enlarge the $[-5,5]^2$ windows of the full triangular planar domains. Region colors track correspondence; rightmost panels show facewise log-condition distortion $d_T$ on the final corrected maps, viewed from the south pole. Polar scores $R_p^d$ follow \eqref{eq:polar_risk} and combine both caps. Distortion uses shared square-root color normalization, with ticks in the original $d_T$ values.}

\label{fig:flash-preconditioning-pipeline}
\end{figure}

Figure~\ref{fig:flash-preconditioning-pipeline} illustrates the intervention on the Max Planck surface. Planar preconditioning is inserted before inverse stereographic projection, after which the same south-pole correction rule is applied to obtain the final spherical map.

\subsection{Geometric control of the affine perturbation family}
\label{subsec:flash-affine-control}

We bound the anisotropic part of \eqref{eq:flash-affine-family} to control the geometric distortion introduced by the representation change.  Write
\begin{equation}
    B=R_\phi\begin{pmatrix}\eta&0\\0&-\eta\end{pmatrix}R_\phi^T.
    \label{eq:flash-B-diag}
\end{equation}
In complex notation, with $b=t_1+\mathrm i t_2$,
\begin{equation}
    w
    =\mathrm e^a\left(
      \cosh\eta\,z
      +\mathrm e^{2\mathrm i\phi}\sinh\eta\,\overline z
    \right)+b.
    \label{eq:flash-affine-complex}
\end{equation}

\begin{figure}[t]
\centering
\includegraphics[width=\linewidth]{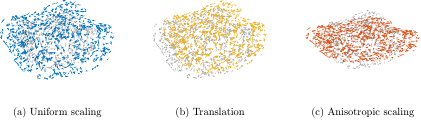}
\caption{\textbf{Geometric interpretation of the affine chart preconditioner.} The three panels illustrate the scale, translation, and anisotropy components of the planar map $\mathcal A_\theta(z)=\exp(a)\exp(B)z+t$ used to generate candidate representations.}
\label{fig:affine-chart-preconditioner}
\end{figure}

Figure~\ref{fig:affine-chart-preconditioner} gives a geometric interpretation of the three principal actions in this family. The scalar parameter $a$ changes the chart scale, $b$ translates the planar realization, and the traceless symmetric matrix $B$ introduces directional anisotropy. These actions are coupled in the completed spherical map through the radial factors of the inverse stereographic lift and the subsequent harmonic correction.

\begin{proposition}[Beltrami coefficient of the affine representation map]
\label{prop:flash-affine-beltrami}
The affine map \eqref{eq:flash-affine-complex} has constant Beltrami coefficient
\begin{equation}
    \mu_{\mathcal A}
    =\mathrm e^{2\mathrm i\phi}\tanh\eta,
    \label{eq:flash-affine-mu}
\end{equation}
and maximal quasiconformal dilation
\begin{equation}
    K(\mathcal A)=\mathrm e^{2|\eta|}.
    \label{eq:flash-affine-K}
\end{equation}
\end{proposition}

\begin{proof}
For $\phi=0$,
$\mathrm e^\eta x+\mathrm i\mathrm e^{-\eta}y
=\cosh\eta\,z+\sinh\eta\,\overline z$. Conjugating by rotations introduces the factor $\mathrm e^{2\mathrm i\phi}$.  Hence $w_z=\mathrm e^a\cosh\eta$ and $w_{\overline z}=\mathrm e^a\mathrm e^{2\mathrm i\phi}\sinh\eta$, which gives \eqref{eq:flash-affine-mu}.  The standard quasiconformal relation \cite{Ahlfors2006} $K=(1+|\mu|)/(1-|\mu|)$ gives \eqref{eq:flash-affine-K}.
\end{proof}

At the raw stereographic stage, the five directions have an explicit first-order action on finite chord lengths.  Let $d_{ij}=z_i-z_j\neq0$, $\widehat d_{ij}=d_{ij}/\|d_{ij}\|_2$, and $q_r=1+\|z_r\|_2^2$. We have the following result:

\begin{proposition}[First-order action on completed edge lengths]
\label{prop:flash-edge-sensitivity}
At the identity representation, let $(a,B,t)
    =\varepsilon(\dot a,\dot B,\dot t)+o(\varepsilon)$. Then
\begin{equation}
    \left.\frac{\mathrm d}{\mathrm d\varepsilon}
    \log\ell_{ij}(\varepsilon)\right|_{\varepsilon=0}
    =
    \dot a+\widehat d_{ij}^T\dot B\widehat d_{ij}
    -
    \frac{z_i^T(\dot aI_2+\dot B)z_i+z_i^T\dot t}{q_i}
    -
    \frac{z_j^T(\dot aI_2+\dot B)z_j+z_j^T\dot t}{q_j},
    \label{eq:flash-edge-sensitivity}
\end{equation}
where $\ell_{ij}(\varepsilon)$ is the chord length after inverse stereographic projection.
\end{proposition}

\begin{proof}
By Proposition~\ref{prop:flash-finite-edge},
\begin{equation}
    \ell_{ij}(\varepsilon)
    =\frac{2\|A_\varepsilon d_{ij}\|_2}
    {\sqrt{q_i(\varepsilon)q_j(\varepsilon)}},
    \qquad
    A_\varepsilon=I_2+\varepsilon(\dot aI_2+\dot B)+o(\varepsilon).
\end{equation}
Differentiating the logarithm gives \eqref{eq:flash-edge-sensitivity}.
\end{proof}

Translation affects the radial denominators while preserving planar edge differences; anisotropy also changes the numerator according to the edge direction. After correction, the dependence is through the transformed anchor data and their harmonic extension, as shown in Theorem~\ref{thm:flash-closed-form}. We therefore evaluate preconditioning candidates on the final corrected map.

\subsection{Bounded parameterization of the affine search}
\label{subsec:implementation}
We optimize over unconstrained coordinates $u\in\mathbb{R}^5$ and map them to the bounded physical parameters of the affine preconditioner by
\begin{equation}
\begin{gathered}
a=2\tanh u_1,\qquad
(b_1,b_2)
=
0.45\frac{\tanh\|v\|}{\|v\|}v,
\qquad
v=(u_2,u_3),\\
t
=
1.25r_0\mathrm{e}^a
(\tanh u_4,\tanh u_5)^T,
\qquad
r_0
=
\left(
\frac{1}{3}
\sum_{i\in T_\infty}|z_i|^2
\right)^{1/2}.
\end{gathered}
\label{eq:parameter_bounds}
\end{equation}
The anisotropy factor is defined by its continuous extension $\tanh\|v\|/\|v\|=1$ at $v=0$. Here $T_\infty$ is the puncture face used in the planar FLASH construction, and $z_i$ are the baseline planar vertices.

This parameterization enforces $|a|<2$ and $\sqrt{b_1^2+b_2^2}<0.45$, while the componentwise translation bounds imply $\|t\|\leq 1.25\sqrt{2}\,r_0\mathrm{e}^a$. The factor $r_0\mathrm{e}^a$ scales the admissible translation with the characteristic size of the planar representation after isotropic scaling. When $r_0$ is numerically zero, we replace it by a median-radius value. All optimization and interpolation are performed in the unconstrained coordinates $u$, while candidate maps are evaluated using the corresponding physical affine parameters $(a,b_1,b_2,t_1,t_2)$.

\subsection{Constrained multistart search}
\label{sec:search}

\subsubsection{Polar objective and admissibility}

In addition to the local log-condition distortion $d_T$, we use the root-mean-square chord-angle distortion (CDI):
\begin{equation}
 c_T(S)
 =
 \left(
 \frac13\sum_{r=1}^3
 (\vartheta_{T,r}^S-\vartheta_{T,r}^V)^2
 \right)^{1/2}.
 \label{eq:cdi}
\end{equation}

For a face $T=(i,j,k)$ with $s_i+s_j+s_k\neq0$, define its centroid direction by $\widehat c_T
 =
 \frac{s_i+s_j+s_k}{\|s_i+s_j+s_k\|}$. For a polar half-angle $\beta$, the north and south face sets are
\begin{equation}
 \mathcal{P}_N(S;\beta)
 =
 \{T:\widehat c_{T,3}\geq\cos\beta\},
 \qquad
 \mathcal{P}_S(S;\beta)
 =
 \{T:\widehat c_{T,3}\leq-\cos\beta\}.
 \label{eq:polar_sets}
\end{equation}
We use $\beta=22.5^\circ$. If either cap contains fewer than five faces, both caps are replaced by the
\[
 k=\min\{|F|,\max(5,\lceil0.05|F|\rceil)\}
\]
faces with largest and smallest centroid heights, respectively. For nonnegative values $x_1,\ldots,x_m$, ordered as $x_{[1]}\geq\cdots\geq x_{[m]}$, define the empirical upper-tail mean
at level $\alpha$ by
\begin{equation}
 \operatorname{UTail}_{\alpha}(x)
 =
 \frac{1}{k_\alpha}
 \sum_{r=1}^{k_\alpha}x_{[r]},
 \qquad
 k_\alpha
 =
 \max\{1,\lceil(1-\alpha)m\rceil\},
 \qquad
 \alpha=0.95.
 \label{eq:upper-tail}
\end{equation}
Thus, $\operatorname{UTail}_{0.95}$ averages the largest $5\%$ of the facewise distortion values. The corresponding polar and global upper-tail log-condition measures are
\begin{align}
 R_p^d(S)
 &=
 \frac12\left[
 \operatorname{UTail}_\alpha(\{d_T:T\in\mathcal{P}_N\})
 +
 \operatorname{UTail}_\alpha(\{d_T:T\in\mathcal{P}_S\})
 \right],
 \label{eq:polar_risk}\\
 R_g^d(S)
 &=
 \operatorname{UTail}_\alpha(\{d_T:T\in F\}),
\end{align}
and replacing $d_T$ by $c_T$ defines $R_p^c$ and $R_g^c$. The polar sets are recomputed for each candidate.

Let $r_p^d,r_p^c,r_g^d,r_g^c$ denote the corresponding normalized upper-tail ratios relative to the original FLASH map $S_0$. For positive baseline upper-tail measures, the final-map objective is
\begin{equation}
 J_p(S)=\max\{r_p^d(S),r_p^c(S)\}.
 \label{eq:polar_minimax}
\end{equation}
A candidate is distortion-feasible if
\begin{equation}
 \begin{gathered}
 r_p^d\leq1,\qquad
 r_p^c\leq1,\qquad
 r_g^d\leq1.01,\qquad
 r_g^c\leq1.01,\\
 \overline d(S)\leq1.02\overline d(S_0),\qquad
 \overline c(S)\leq1.02\overline c(S_0),
 \end{gathered}
 \label{eq:feasible_set}
\end{equation}
where the bars denote arithmetic means over all faces. Thus the objective targets polar upper-tail distortion while the global upper-tail measures and mean distortions prevent deterioration away from the poles. We additionally require $|\log R_{\mathrm{NS}}(Z)|\leq\log(1.10)$, where
\begin{equation}
 R_{\mathrm{NS}}(Z)=
 \frac{\ell(T_N;Z)}{\ell(T_S;W(Z))},
 \qquad W(Z)_i=\frac{z_i}{\|z_i\|^2}.
\end{equation}
Here $\ell$ is the mean triangle edge length, and $T_N, T_S$ are the puncture and inner triangles selected by the original FLASH and held fixed during the search. This constraint is checked on the transformed planar coordinates before south-pole correction.

We additionally require every accepted candidate to define a valid spherical realization: vertices must be finite and lie on the unit sphere, all faces must remain nondegenerate with the baseline orientation, and the map must have the same signed spherical coverage as the baseline. For reference, the signed solid angle of
$T=(i,j,k)$ is \cite{VanOosteromStrackee1983}
\begin{equation}
 \omega_T
 =
 2\operatorname{atan2}
 \left(
 \det[s_i,s_j,s_k],\,
 1+s_i\cdot s_j+s_j\cdot s_k+s_k\cdot s_i
 \right).
 \label{eq:solid_angle}
\end{equation}

\subsubsection{Candidate generation and selection}

Evaluating every trial with the full south-pole correction would make the search unnecessarily expensive. We therefore use the raw inverse-stereographic map only to generate a small set of candidates, and evaluate the final objective and feasibility only after applying the unchanged FLASH correction.

A raw evaluation applies $\mathcal{A}_\theta$ followed by $\Pi_N^{-1}$, without the south-pole solve. Using the corresponding baseline-normalized raw risks, candidate generation minimizes
\begin{equation}
\resizebox{0.9\linewidth}{!}{%
 $J_{p,\mathrm{raw}}
 ={}\max\{r_{p,\mathrm{raw}}^d,r_{p,\mathrm{raw}}^c\}
 +\left[
 \max\{r_{g,\mathrm{raw}}^d,r_{g,\mathrm{raw}}^c\}-1.01
 \right]_+^2
 +10^5\left[
 |\log R_{\mathrm{NS}}(Z)|-\log(1.10)
 \right]_+^2,$
 }
\label{eq:raw_objective}
\end{equation}
where $[x]_+=\max\{x,0\}$. Invalid raw maps are rejected.

Algorithm~\ref{alg:preconditioning} uses similarity and fixed-scale warm starts, followed by two full-affine searches. Five candidates are sampled between the similarity solution and the selected affine endpoint. Optimization and interpolation use the unconstrained coordinates of Section~\ref{subsec:implementation}, decoded through \eqref{eq:parameter_bounds}. The evaluation budgets are upper-limit settings; convergence or stagnation may terminate a search earlier.

\begin{algorithm}[tbp]
\caption{Pole-aware planar preconditioning}
\label{alg:preconditioning}
\begin{algorithmic}[1]
\Require Source mesh, puncture face, and the original FLASH map.
\Ensure A feasible corrected spherical map with the smallest evaluated $J_p$.

\State Compute the balanced coordinates $Z_0$ and the original FLASH map $S_0$ at the same puncture face.

\State Record the baseline risks and retain $S_0$ as a candidate.

\State Starting from zero search coordinates, compute a similarity warm start $u_{\mathrm{sim}}$ with $b_1=b_2=0$ and a fixed-scale four-parameter warm start $u_{\mathrm{res}}$, using raw objective evaluation budgets of 40 and 70, respectively.

\State Optimize all five parameters from each warm start with a budget of 70 per run. Retain the endpoint $u_{\mathrm{aff}}$ with the smaller raw objective.

\For{$\lambda\in\{0,\frac14,\frac12,\frac34,1\}$}
    \State Form
    $u_\lambda=u_{\mathrm{sim}}
    +\lambda(u_{\mathrm{aff}}-u_{\mathrm{sim}})$
    and decode the corresponding planar map.
    \State Reject nonfinite raw maps and candidates violating the
    north--south constraint; otherwise apply the original south-pole
    correction and test final-map admissibility.
\EndFor

\State \Return the map with the smallest objective
\eqref{eq:polar_minimax} among all feasible corrected candidates,
including $S_0$.
\end{algorithmic}
\end{algorithm}

Because the original map $S_0$ is retained as a baseline candidate, the returned map satisfies $J_p\leq1$ whenever $S_0$ is admissible and the baseline risks are positive. This safeguard applies to the aggregate criteria in \eqref{eq:feasible_set} and does not imply facewise improvement.

\paragraph{Computational cost.}
Each raw evaluation requires $O(n+|F|)$ geometric work, while full sorting for the tail statistics gives the conservative bound $O(n+|F|\log|F|)$. Hence, with $E$ raw evaluations and at most five additional south-pole corrections, the additional work beyond the reused planar construction is
\begin{equation}
 O\bigl(E(n+|F|\log|F|)\bigr)+qC_{\mathrm{south}},
 \qquad q\leq5.
 \label{eq:cost}
\end{equation}
Candidate evaluation retains the original FLASH south-pole solver. The equivalent harmonic realization of the final selected output is performed separately and is not included in this search cost.

\section{Experimental results}
\label{sec:experiment-design}

In this section, we present our numerical study to address three questions. First, we verify the discrete identities underlying the FLASH analysis and the equivalent harmonic realization of the south-pole correction. Second, we evaluate whether the proposed planar preconditioning improves the final corrected FLASH map and isolate the mechanisms responsible for the improvement. Third, we examine how representation perturbations propagate through iterative spherical algorithms and whether the preconditioning effect remains visible in downstream computations. Genus-zero closed meshes from the TU/e collection~\cite{tuealga_meshes2026} are used for our experiments. All numerical experiments were performed on a machine equipped with a 14-core Apple M4 Pro CPU (10 performance and 4 efficiency cores) and 48~GB of unified memory.

\subsection{Perturbation and measurement protocol}
\label{subsec:experiment-common}

For the local sensitivity experiments, let $\xi$ be a normalized parameter or state direction and consider $\theta(\varepsilon)=\varepsilon\xi$. Each method is evaluated on a logarithmic sequence of small positive and negative perturbation magnitudes. Central differences are used whenever the $+\varepsilon$ and $-\varepsilon$ runs remain in the same computational regime.

For the state at solver step $k$, define
\begin{equation}
    \Delta_k(\varepsilon)
    :=
    \left\|X^{(k)}_{\varepsilon}-X^{(k)}_0\right\|,
    \qquad
    A_k(\varepsilon)
    :=
    \frac{\Delta_k(\varepsilon)}{\Delta_0(\varepsilon)}.
    \label{eq:exp-trajectory-amplification}
\end{equation}
When two spherical outputs differ by a physically irrelevant global rotation, the same prescribed $\mathrm{SO}(3)$ alignment is applied before evaluating the norm. State norms are Euclidean norms after vectorization, equivalently Frobenius norms for vertex-coordinate matrices; distinct state components are treated separately.

For a predicted tangent $\dot X^{(k)}$, the relative central-difference error is
\begin{equation}
    E_k^{\mathrm{FD}}(\varepsilon)
    :=
    \frac{
    \left\|
    \dfrac{X^{(k)}_{+\varepsilon}-X^{(k)}_{-\varepsilon}}
          {2\varepsilon}
    -\dot X^{(k)}
    \right\|
    }{
    \max\{\|\dot X^{(k)}\|,\varepsilon_{\mathrm{mach}}\}
    }.
    \label{eq:exp-fd-error}
\end{equation}
The linear-response window is the range over which the central-difference estimate remains stable as $\varepsilon$ decreases without a change of computational branch. For the iterative experiments, we report terminal and peak amplification, together with changes in convergence or geometric validity when they occur.

\subsection{FLASH structural validation}
\label{subsec:experiment-flash-structure}

We first verify the discrete structure derived in Section~\ref{sec:flash-exact}. These tests separate the accuracy of the analytic identities from the subsequent optimization experiments.

\begin{figure}[t]
\centering
\includegraphics[width=0.88\linewidth]{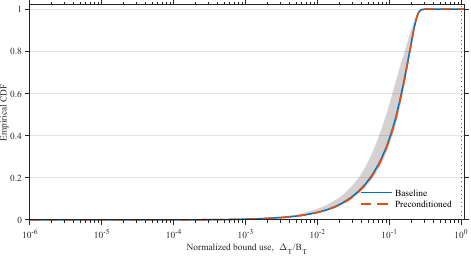}
\caption{\textbf{Empirical distribution of normalized utilization of the inverse-stereographic finite-triangle conditioning bound on eligible faces.} Baseline and preconditioned representations use the same source-face correspondence; the vertical reference at one denotes saturation of the sufficient bound.}
\label{fig:flash-raw-bound}
\end{figure}

For the inverse-stereographic stage, the observed change in finite-triangle conditioning is compared with the sufficient bound in Theorem~\ref{thm:flash-radial-conditioning} on faces satisfying its hypothesis. Figure~\ref{fig:flash-raw-bound} reports the empirical distribution of normalized bound utilization for the baseline and preconditioned planar representations.

\begin{figure}[t!]
\centering
\includegraphics[width=\linewidth]{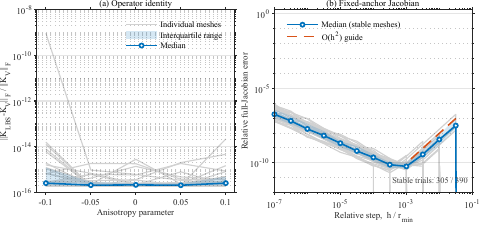}
\caption{
\textbf{Numerical validation of the exact FLASH structure.}  (a) Relative discrepancy between the unmodified LBS stiffness matrix and the intrinsic source Dirichlet matrix under controlled affine perturbations. (b) Relative error of the analytic complete-map Jacobian against central finite differences within a fixed-anchor regime. The dashed guide indicates second-order finite-difference behavior.
}
\label{fig:flash-operator-jacobian}
\end{figure}

We next test the exact LBS/source-Dirichlet reduction. For each tested planar realization, the assembled unmodified LBS stiffness matrix is compared with the intrinsic source Dirichlet matrix using
\begin{equation}
    E_K
    :=
    \frac{\|K_{\mathrm{LBS}}-K_V\|_F}{\|K_V\|_F}.
    \label{eq:exp-flash-operator-error}
\end{equation}
Figure~\ref{fig:flash-operator-jacobian}(a) shows discrepancies at the level of floating-point precision across the tested affine perturbations, consistent with the exact operator identity.

The affine perturbation experiment also provides a numerical check of the complete fixed-anchor Jacobian. The directional derivative in \eqref{eq:flash-jacobian} is compared with a central finite difference of the final corrected spherical map. Among the 390 finite-difference trials, 305 remained within the stable fixed-anchor analytic regime. Within this stable subset, Figure~\ref{fig:flash-operator-jacobian}(b) exhibits the expected second-order finite-difference regime before finite-precision and finite-amplitude effects dominate.

\subsubsection{Numerical equivalence of the harmonic realization}
\label{subsec:experiment-flash-equivalence}

The matrix identity also gives an alternative implementation of the south-pole correction. We therefore apply the original LBS correction and the source-Dirichlet harmonic extension to identical anchor data and compare the resulting spherical maps.

\begin{table}[t]
\centering\small
\caption{Numerical equivalence of the original LBS south-pole correction and the direct source-harmonic realization in the mechanism-ablation experiment. The two reported contexts use the same 30-mesh cohort and identical south-correction inputs within each context. The maximum vertex discrepancy is the maximum Euclidean vertex difference over a context; the RMS column is the median per-mesh vertex RMS difference. Speedup is original correction time divided by harmonic-extension time.}
\label{tab:south-equivalence}
\setlength{\tabcolsep}{4pt}
\resizebox{\linewidth}{!}{%
\begin{tabular}{@{}lrrrr@{}}
\toprule
Context & $N$ & Max vertex diff. & Median RMS diff. & Median speedup \\
\midrule
Original balance, no preconditioning & 30 & $1.11\times10^{-13}$ & $1.15\times10^{-14}$ & $1.62\times$ \\
Original balance, preconditioned & 30 & $1.13\times10^{-13}$ & $1.26\times10^{-14}$ & $1.64\times$ \\
\midrule
Selected original/preconditioned contexts & 60 & $1.13\times10^{-13}$ & $1.19\times10^{-14}$ & $1.63\times$ \\
\bottomrule
\end{tabular}}
\par\vspace{0.4em}
\footnotesize Retry/fixed-anchor/validity mismatches: 0/0/0 across all 60 paired corrections; the harmonic realization is faster in 60/60 runs. 
\end{table}

Table~\ref{tab:south-equivalence} compares the original LBS correction with the source-Dirichlet harmonic realization on identical corrected inputs. The two implementations agree to numerical precision, with no observed differences in retry decisions, anchor sets, or geometric validity. The reported speedup concerns the correction stage only and excludes the planar preconditioner search. The harmonic form therefore provides an equivalent and faster realization of the same correction.

\subsection{FLASH preconditioning: final quality and mechanism}
\label{subsec:experiment-flash-preconditioning}

We next evaluate the bounded five-parameter affine preconditioner on the complete corrected FLASH map. The resulting map is referred to as the preconditioned FLASH map. It consists of the original north--south balancing, the bounded five-parameter planar preconditioner, inverse stereographic completion, and the final south-pole correction. The original FLASH and the preconditioned maps use the same puncture face, north--south balancing procedure, and south-pole correction rule; the preconditioned map uses the source-harmonic realization for the final selected output, whereas the original LBS realization is retained for the original baseline and for candidate evaluation during the search. The original FLASH map is retained as the baseline candidate throughout the search.

\subsubsection{Geometric effect of planar preconditioning}

\begin{table}[htbp]
\centering\small
\caption{Cohort-level effect of the proposed planar preconditioning on 30 genus-0 closed surfaces. Distortion gains are defined as $100(1-R_{\mathrm{pre}}/R_{\mathrm{off}})\%$, so positive values indicate lower distortion. The angle row reports the paired change in the lower $5\%$ polar minimum-angle quantile. Confidence intervals are percentile bootstrap intervals for the paired median~\cite{EfronTibshirani1993} (10000 mesh resamples). The runtime row reports only the preconditioner stage, with all south-pole correction work excluded.}
\label{tab:main-results-updated}
\setlength{\tabcolsep}{4.5pt}
\begin{tabular}{@{}lrrrr@{}}
\toprule
Metric & Median & IQR & 95\% CI & Favorable \\
\midrule
Polar $d$ gain (\%) & 10.17 & 10.35 & [6.29, 14.34] & 100\% \\
Polar CDI gain (\%) & 10.97 & 10.58 & [6.76, 13.93] & 100\% \\
Global $d$ gain (\%) & 1.49 & 4.21 & [0.22, 2.45] & 93\% \\
Global CDI gain (\%) & 1.81 & 4.82 & [0.20, 2.94] & 93\% \\
Polar minimum-angle gain ($^\circ$) & 0.080 & 0.357 & [0.007, 0.234] & 87\% \\
\bottomrule
\end{tabular}
\par\vspace{0.8em}
\begin{tabular}{@{}lrr@{}}
\toprule
Preconditioner stage & Median (s) & IQR (s) \\
\midrule
Planar preconditioner only & 1.480 & 0.596 \\
\bottomrule
\end{tabular}
\par\vspace{0.35em}
\footnotesize This stage includes the original north-stage construction and the constrained planar search, but excludes both candidate south-correction calls and the final south correction.
\end{table}

Table~\ref{tab:main-results-updated} reports the geometric effect of the selected planar preconditioner and the computational cost of the preconditioning stage, with south-pole correction work excluded. The numerical equivalence and correction-only computational cost of the original LBS correction and the source-Dirichlet harmonic realization are reported separately in Table~\ref{tab:south-equivalence}.

\begin{figure}[t!]
\centering
\includegraphics[width=0.95\linewidth]{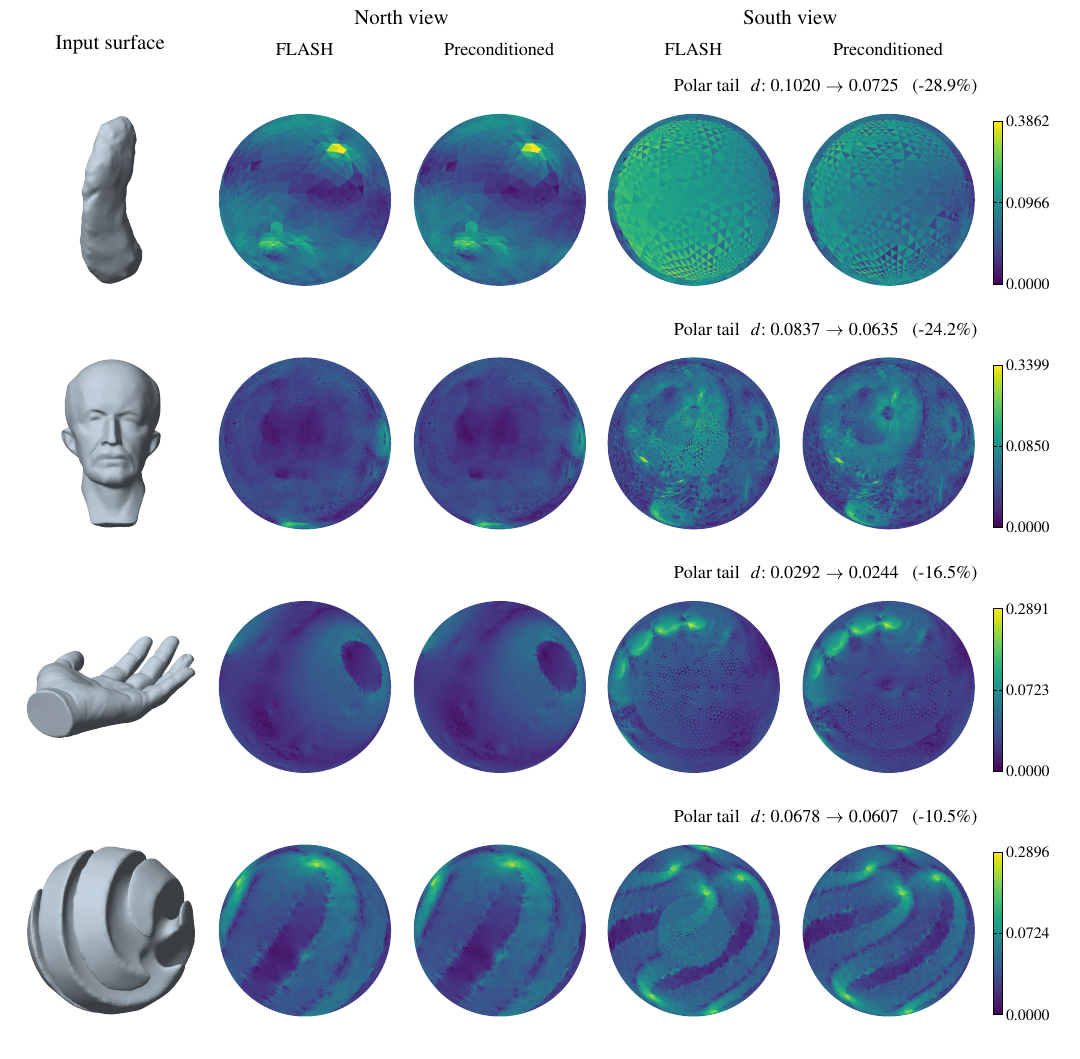}
\caption{
\textbf{Representative comparisons of the original FLASH~\cite{ChoiLamLui2015FLASH} and the preconditioned spherical parameterization results for four genus-0 closed surfaces.}
Rows show Hippocampus, Max Planck, Camille hand, and Twisted ball. Each row contains the input surface, followed by paired north views and paired south views of the final corrected spherical maps. Colors represent facewise log-condition distortion $d_T$. The four spherical views within each row share a color scale; ranges differ between rows. Colors use a square-root scale, with colorbar ticks labeled in the original $d_T$ values. Annotations report the polar upper-tail measure $R_p^d$ defined in \eqref{eq:polar_risk}, before and after preconditioning, together with its percentage reduction.
}
\label{fig:flash-spatial-example}
\end{figure}

The principal effect is concentrated near the poles. The two polar tail-distortion measures decrease consistently across the cohort, whereas the corresponding global reductions are more modest. This separation is consistent with the design of the objective: the search targets the regions most exposed to chart-induced distortion while constraining deterioration elsewhere.
Figure~\ref{fig:flash-spatial-example} compares the final corrected maps on four representative surfaces. For each surface, the original FLASH and preconditioned parameterization results are shown from the north and south views using the same color scale. The annotated polar tail values summarize the reduction in $R_p^d$, complementing the cohort-level results in Table~\ref{tab:main-results-updated}.

\begin{figure}[t!]
\centering
\includegraphics[width=\linewidth]{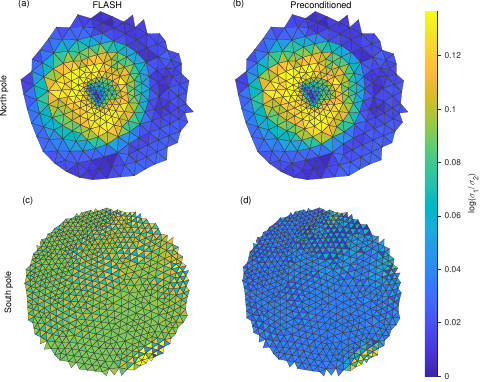}
\caption{\textbf{Local polar comparison of original FLASH (left) and the preconditioned map (right), with north and south regions shown separately.} Within each row, the two maps use the same displayed scale, highlighting the reduction of localized polar distortion under preconditioning.}
\label{fig:polar_patches}
\end{figure}

Figure~\ref{fig:polar_patches} focuses on enlarged neighborhoods around the north and south poles. At this local scale, the preconditioned map suppresses several of the most strongly distorted faces and produces a more regular arrangement of neighboring elements. These enlarged views complement the whole-sphere comparisons in Figure~\ref{fig:flash-spatial-example} by showing the local changes in polar distortion and element arrangement.

\subsubsection{Effect of the south-pole correction}
\label{subsec:experiment-flash-stagewise}

To isolate how the planar intervention propagates through the final FLASH stage, we evaluate the same preconditioned representation both immediately after inverse stereographic completion and after the unchanged south-pole correction (see Table~\ref{tab:stage-amplification-updated}).

\begin{table}[t]
\centering\small
\caption{Mechanism isolation on the 30-mesh cohort. The raw columns measure the effect before the final south-pole correction; the corrected columns measure the same paired perturbation after the correction. Positive values indicate reduced distortion.}
\label{tab:stage-amplification-updated}
\setlength{\tabcolsep}{5pt}
\resizebox{\linewidth}{!}{%
\begin{tabular}{@{}lrrr@{}}
\toprule
Metric & Raw gain (median [IQR]) & Corrected gain (median [IQR]) & Corrected favorable \\
\midrule
Polar $d$ & 2.56\% [2.77] & 10.17\% [10.35] & 100\% \\
Polar CDI & 2.58\% [2.92] & 10.97\% [10.58] & 100\% \\
\bottomrule
\end{tabular}}
\end{table}

The preconditioned representation reduces the median raw polar $d$ and polar CDI by $2.56\%$ and $2.58\%$, respectively. After the unchanged south-pole correction, the corresponding median reductions increase to $10.17\%$ and $10.97\%$. The correction therefore propagates the modified anchor data in a way that strengthens the polar improvement predicted at the inverse-stereographic stage. This behavior is consistent with the fixed-anchor factorization in Section~\ref{sec:flash-exact}. The result also demonstrates the computational value of upstream representation control: a low-dimensional planar adjustment produces a larger improvement in the final corrected map while preserving the existing correction procedure. Planar preconditioning therefore supplies an efficient mechanism for transferring improved chart geometry to the final spherical parameterization.

\begin{figure}[t]
\centering
\includegraphics[width=0.92\linewidth]{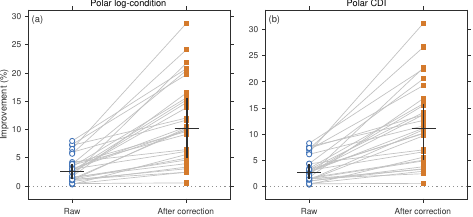}
\caption{\textbf{Stagewise effect of the same planar preconditioning perturbation on the 30-mesh cohort.} Each thin segment connects the raw and corrected measurements for one mesh; thick segments and ticks summarize the interquartile range and median. Left: polar log-condition distortion. Right: polar CDI.}
\label{fig:flash-stagewise-effect}
\end{figure}

\begin{table}[t!]
\centering
\caption{Degree-of-freedom ablation on the 30-mesh cohort.
The 3-DOF similarity family, 4-DOF residual family with frozen isotropic scale, and full 5-DOF affine family are compared. Distortion gains are paired median improvements relative to original FLASH; positive values indicate reduced polar distortion.}
\label{tab:dof_ablation}
\begin{tabular}{lccc}
\toprule
Configuration & DOF & Polar $d$ gain (\%) & Polar CDI gain (\%) \\
\midrule
Similarity & 3 & 0.17 & 0.16 \\
Residual (fixed scale) & 4 & 1.59 & 1.55 \\
Full affine & 5 & 10.99 & 11.88 \\
\bottomrule
\end{tabular}
\end{table}

\subsubsection{Parameter-family ablation}
\label{subsec:experiment-flash-ablation}

We finally examine which components of the affine representation are needed to obtain the observed improvement. All configurations start from the same originally balanced FLASH coordinates and use the same geometric constraints and final-map evaluation protocol. We compare a three-degree-of-freedom similarity family, a four-degree-of-freedom family with isotropic scale fixed, and the full five-degree-of-freedom affine family. The search budgets differ: 250 for similarity, 75 for the fixed-scale family, and warm-start budgets of 50 and 75 followed by two full-affine searches with budgets of 75 each. Thus, the comparison does not isolate parameter count under equal computational budgets. See Table~\ref{tab:dof_ablation} for the results.

The full affine configuration achieves the largest reductions in both polar distortion measures, with median gains of 10.99\% in polar $d$ and 11.88\% in polar CDI. These results support the effectiveness of the five-parameter affine model with the multistart search used in this experiment.

\subsection{Representation sensitivity in iterative spherical parameterization}
The FLASH analysis above exploits method-specific structure. We next apply the local perturbation protocol to cMCF and SDEM. Classical analysis of linear evolution shows that intermediate amplification can differ substantially from asymptotic behavior~\cite{TrefethenEmbree2005}. Here, we examine transient and terminal responses along the state-dependent parameterization trajectories introduced in Section~\ref{sec:rs-framework}.

\subsubsection{cMCF}

The cMCF experiment follows the geometric evolution described in Section~\ref{subsec:cmcf-realization}. In addition to the native convergence stopping rule, each perturbed trajectory is evaluated at the stopping index of its unperturbed baseline. This matched-horizon control separates trajectory sensitivity from changes in the stopping index.

\begin{table}[t!]
\centering
\caption{Representation-sensitivity statistics for cMCF under native convergence stopping and the baseline-matched horizon control. For mesh $i$, $K_i^0$ is the stopping iteration of the unperturbed baseline and $K_{i,\epsilon}$ is the first tolerance iteration of the perturbed run. The native terminal amplification is evaluated at $K_{i,\epsilon}$, whereas $A_{\mathrm{match}}$ evaluates the perturbed and baseline trajectories at the common mesh-specific horizon $K_i^0$. Amplification entries are median $[Q_{25},Q_{75}]$, pooling the four perturbation directions and both signs.}
\label{tab:cmcf_sensitivity}
{\footnotesize
\setlength{\tabcolsep}{4.2pt}
\begin{tabular}{@{}c r c c c r@{}}
\hline
$|\epsilon|$ & Runs & $A_{\mathrm{term}}$ & $A_{\mathrm{match}}$ & $\max_{k\le K_i^0} A_k$ & $K_{i,\epsilon}\neq K_i^0$ \\
\hline
$10^{-3}$            & 240 & $0.353\,[0.180,\,0.640]$ & $0.348\,[0.180,\,0.631]$ & $1.365\,[1.061,\,2.136]$ & $1\;(0.4\%)$ \\
$3\times10^{-3}$     & 240 & $0.352\,[0.180,\,0.649]$ & $0.347\,[0.180,\,0.632]$ & $1.365\,[1.061,\,2.137]$ & $3\;(1.3\%)$ \\
$10^{-2}$            & 240 & $0.350\,[0.180,\,0.640]$ & $0.347\,[0.180,\,0.634]$ & $1.365\,[1.061,\,2.141]$ & $5\;(2.1\%)$ \\
$3\times10^{-2}$     & 240 & $0.351\,[0.182,\,0.640]$ & $0.345\,[0.180,\,0.639]$ & $1.366\,[1.063,\,2.144]$ & $18\;(7.5\%)$ \\
\hline
All                   & 960 & $0.352\,[0.180,\,0.641]$ & $0.346\,[0.180,\,0.639]$ & $1.366\,[1.063,\,2.144]$ & $27\;(2.8\%)$ \\
\hline
\end{tabular}
}
\end{table}

As shown in Table~\ref{tab:cmcf_sensitivity}, the native and matched-horizon terminal summaries are close across all perturbation magnitudes, indicating that changes in the stopping index have a limited effect on the aggregate terminal amplification in this experiment. Small initial perturbations are typically attenuated by the terminal state, with median terminal amplification remaining below one. However, the trajectory-level response reveals transient amplification: the median of $\max_{k\le K_i^0}A_k$ is approximately $1.37$, and its upper quartile exceeds $2.1$. Thus, terminal attenuation does not fully describe the numerical response of cMCF, because perturbations can grow substantially during the geometric evolution before being reduced at later steps.

Changes in the stopping index remain uncommon for small perturbations but become more frequent as the perturbation magnitude increases, reaching $7.5\%$ of runs at $|\epsilon|=3\times10^{-2}$. This pattern is consistent with the product of state-dependent step Jacobians derived for cMCF: intermediate amplification can coexist with terminal attenuation, while sufficiently large perturbations can modify the convergence trajectory and stopping decision. The matched-horizon comparison provides a direct way to separate propagation along a common number of steps from effects associated with the stopping rule.

These results suggest a sensitivity-aware route for improving cMCF initialization. Initial representations could be evaluated using both terminal amplification and trajectory-level quantities such as peak amplification and stopping-index stability. Future work could optimize representation-preserving initial transformations that reduce transient amplification, maintain a larger convergence margin, and improve the reliability of the stopping decision. This would extend the representation-control strategy developed for FLASH to iterative geometric evolution, where the quality of an initialization can be assessed through its full trajectory and its downstream convergence behavior.

\subsubsection{SDEM}

The SDEM sensitivity experiment uses the coupled state described in Section~\ref{subsec:sdem-realization}. Only meshes whose unperturbed runs reach the prescribed tolerance and produce valid spherical maps are included in the local sensitivity analysis. Perturbed runs that lose validity or fail to converge are classified as unstable and excluded from the amplification statistics. Table~\ref{tab:sdem_sensitivity} shows the results.

\begin{table}[htbp]
\centering
\caption{SDEM representation-sensitivity statistics after baseline screening. Only meshes whose unperturbed SDEM run both reached the prescribed tolerance and produced a geometrically valid spherical map are eligible for the local sensitivity analysis. A perturbed run is classified as stable when it also satisfies both conditions. Amplification statistics are computed only over stable runs. Entries are median $[Q_{25},Q_{75}]$; the last two columns report counts and percentages within the stable subset.}
\label{tab:sdem_sensitivity}
{\footnotesize
\setlength{\tabcolsep}{4.0pt}
\begin{tabular}{@{}c c c c c c@{}}
\hline
$|\epsilon|$ & Stable / eligible & $A_{\mathrm{term}}$ & $\max_k A_k$ & $A_{\mathrm{term}}>1$ & $\max_k A_k>1$ \\
\hline
$10^{-3}$        & $175/192$ & $0.255\,[0.171,\,0.605]$ & $1.000\,[1.000,\,12.339]$ & $38\;(21.7\%)$ & $47\;(26.9\%)$ \\
$3\times10^{-3}$ & $176/192$ & $0.253\,[0.171,\,0.564]$ & $1.000\,[1.000,\,8.382]$  & $25\;(14.2\%)$ & $53\;(30.1\%)$ \\
$10^{-2}$        & $180/192$ & $0.256\,[0.171,\,0.525]$ & $1.000\,[1.000,\,3.828]$  & $14\;(7.8\%)$  & $67\;(37.2\%)$ \\
$3\times10^{-2}$ & $175/192$ & $0.256\,[0.169,\,0.445]$ & $1.000\,[1.000,\,1.596]$  & $9\;(5.1\%)$   & $56\;(32.0\%)$ \\
\hline
All               & $706/768$ & $0.255\,[0.171,\,0.518]$ & $1.000\,[1.000,\,2.874]$ & $86\;(12.2\%)$ & $223\;(31.6\%)$ \\
\hline
\end{tabular}
}
\end{table}

Within the stable subset, terminal perturbations are again typically attenuated, but the peak response is considerably more heterogeneous across meshes and perturbation directions. Some trajectories undergo substantial transient amplification, and a smaller subset remains amplified at termination. This distinction indicates that terminal attenuation alone does not fully characterize the numerical sensitivity of SDEM: intermediate amplification may still affect solver robustness, geometric validity, or the quality of downstream states before the final iteration. The broader variation is consistent with perturbation propagation through the coupled nonlinear state, in which the spherical map and auxiliary variables jointly determine the subsequent updates. These observations suggest that the sensitivity behavior of SDEM can guide the construction of initial representations for subsequent computation. Transient amplification, convergence changes, and geometric-validity loss provide measurable signals of the numerical quality of an initial state. Future work could incorporate these signals into a joint representation-optimization objective that penalizes peak perturbation amplification, reduces convergence risk, and preserves geometric validity while maintaining the desired terminal parameterization quality. Such an approach would extend the sensitivity-aware strategy developed for FLASH to iterative spherical parameterization and could produce initial representations with improved numerical stability and downstream geometric quality.

Runs that leave the stable regime indicate a reduced stability margin of the corresponding initial representation. Their occurrence can therefore be used to identify unfavorable initial states and to guide representation adjustment, candidate rejection, or adaptive initialization in future SDEM-based computations.

\subsection{Downstream consequences}
\label{sec:downstream}

We finally ask whether the change in FLASH initialization remains visible after subsequent processing. All paired downstream experiments use identical geometry and downstream settings.

\subsubsection{SDEM terminal quality and validity}
\label{sec:downstream-sdem}

Both initializations undergo the same M\"obius area correction before SDEM~\cite{LyuLuiChoi2024SDEM} and are then evolved with identical solver settings. The analysis is restricted to paired runs for which the original initialization reaches the prescribed stopping criterion.

\begin{table}[t]
\centering\small
\caption{Terminal SDEM quality and validity on the same 26 paired runs. Paired change is the median per-mesh percentage change $100(R_{\mathrm{pre}}/R_{\mathrm{off}}-1)\%$. Negative distortion changes are improvements.}
\label{tab:downstream-sdem-quality-updated}
\setlength{\tabcolsep}{4.5pt}
\begin{tabular}{@{}lrrr@{}}
\toprule
Metric & Original & Preconditioned & Paired change \\
\midrule
Final density CV & 0.00560 & 0.00572 & $+0.44\%$ \\
Fixed-polar $d$ & 1.22278 & 1.21986 & $-0.17\%$ \\
Fixed-polar CDI & 0.69013 & 0.68505 & $-0.11\%$ \\
Global $d$ & 1.58995 & 1.58965 & $+0.01\%$ \\
Global CDI & 0.89179 & 0.89213 & $-0.00\%$ \\
\midrule
Residual flips, total & 10 & 5 & -- \\
Meshes with residual flips & 3 & 3 & -- \\
\bottomrule
\end{tabular}
\end{table}

As shown in Table~\ref{tab:downstream-sdem-quality-updated}, the terminal distortion summaries remain comparable between the two initializations, while the preconditioned FLASH initialization reduces the total number of residual flips from 10 to 5 in the paired SDEM runs. This result provides evidence that planar preconditioning improves downstream numerical behavior in addition to improving the initial spherical representation. By supplying a more favorable initialization, the preconditioner reduces residual folding events while preserving comparable terminal distortion. These findings support representation preconditioning as an upstream strategy for transferring numerical and geometric benefits from the initial parameterization to subsequent spherical processing.

\subsubsection{Fixed-template remeshing}
\label{subsec:downstream-remeshing}

To test whether the chart intervention remains visible in a subsequent sampling and remeshing task~\cite{PraunHoppe2003,ChoiManLui2016QC,ChoiHoLui2016SpherePointCloud}, we pull both the original and preconditioned spherical maps back to the same fixed icosphere templates at three resolutions: 642, 2562, and 10242 vertices. We measure upper-tail template-to-surface log-condition distortion, edge-length variation, and the lower-tail minimum angle (see Table~\ref{tab:remeshing_cohort}).

\begin{table}[t]
\centering
\small
\caption{Advantages of the preconditioned map in fixed-template remeshing over the 30-mesh cohort. Values report the median relative change with respect to the original FLASH; negative values indicate improvement for risk metrics, while positive values indicate improvement for the minimum-angle metric. The favorable count reports the number of meshes improved out of 30.}
\label{tab:remeshing_cohort}

\resizebox{\linewidth}{!}{%
\begin{tabular}{lccc}
\toprule
Metric
& 642 vertices
& 2,562 vertices
& 10,242 vertices \\
\midrule
Template-to-surface LogCond upper-tail mean (top $5\%$)
& $-0.16\%$ (16/30)
& $-0.37\%$ (21/30)
& $-0.77\%$ (24/30) \\

Edge-length CV
& $-0.07\%$ (16/30)
& $-0.18\%$ (20/30)
& $-0.16\%$ (21/30) \\

Minimum-angle $Q_{05}$
& $+0.06\%$ (19/30)
& $+0.06\%$ (24/30)
& $+0.10\%$ (27/30) \\

\bottomrule
\end{tabular}
}
\end{table}

It can be observed that the benefit of the preconditioned maps persists in the downstream remeshing task and becomes increasingly consistent as the template resolution increases. In particular, improvements in template-to-surface conditioning and lower-tail element quality are observed on an increasing fraction of meshes at finer resolutions. This resolution-dependent trend shows that the reduction in chart-induced distortion persists through the subsequent surface-sampling operation. The remeshing results therefore provide downstream evidence that pole-aware preconditioning transfers local-conditioning benefits from the spherical parameterization to later geometric processing, with the clearest gains emerging on finer templates.

\section{Discussion}
\label{sec:discussion}
\subsection{Principal findings and implications}

One principal result of our work is that computational representation can be treated as a design variable in closed-surface parameterization, with the present analysis and optimization developed for genus-0 spherical parameterization. For FLASH, the exact LBS/source-Dirichlet identity reduces the unmodified south-pole correction to a source-mesh harmonic extension. On a fixed-anchor branch, the resulting factorization identifies the three stages through which planar perturbations reach the corrected spherical map: chart transition, harmonic propagation, and inverse stereographic projection. This structure gives the planar preconditioner a direct mathematical basis and explains why its effect is evaluated on the complete corrected map. The cohort experiments show that a low-dimensional planar intervention can produce a substantial geometric improvement. The stagewise experiment further shows that the correction stage acts as an active propagation mechanism for the modified anchor data. The source-harmonic realization preserves the original correction to numerical precision and reduces correction-stage runtime, giving a practical implementation of the same mathematical operator.



The cMCF and SDEM mechanism studies provide a complementary view of iterative sensitivity. Terminal attenuation frequently coexists with transient amplification, and perturbations can alter stopping decisions or geometric validity along the trajectory. Peak amplification, stopping-index stability, convergence margin, and geometric validity therefore provide useful quantities for future initialization design. Together, the FLASH and iterative experiments support representation selection as an algorithmic design variable that affects both the parameterization itself and the computation that follows.

\subsection{Limitations}

The current preconditioner uses a global affine transformation, which couples adjustments across the chart and limits flexibility for spatially heterogeneous distortion. Its exact FLASH sensitivity description is also branch-specific: anchor switching requires piecewise treatment, and modifying the Beltrami coefficients may change the source-Dirichlet operator equivalence underlying the factorization. These conditions define the scope of the present analysis and optimization. For cMCF and SDEM, the work currently reaches the exploration of perturbation-propagation mechanisms; incorporating those sensitivities into the optimization of initial states or iterative updates remains a further methodological step.

\subsection{Future works}

One future direction is to turn the measured sensitivities into explicit representation-control objectives. For FLASH, the fixed-anchor Jacobian can guide localized chart adjustments, while anchor-gap and geometric-validity margins can be incorporated into branch-aware candidate selection. A spatially varying preconditioner could extend the current global affine family and target distortion patterns that cannot be corrected by a single scale, translation, and anisotropy transformation. For cMCF and SDEM, initial representations can be evaluated using trajectory-level objectives that combine terminal amplification, peak amplification, stopping-index stability, convergence margin, and geometric validity. These quantities can support optimization or adaptive selection of initial states before the iterative evolution begins. The SDEM coupled-state formulation also provides a natural setting for studying how perturbations in auxiliary variables influence the final spherical map.

More broadly, the representation-sensitivity viewpoint may be extended to parameterizations of other closed-surface topologies. Existing work on genus-one and higher-genus surface parameterization provides natural settings for investigating such extensions~\cite{YaoChoi2026TDEM,GuYau2003}. Such extensions would require representation changes compatible with the chosen target surface. The present spherical analysis provides a concrete starting point for these investigations.


\section{Conclusion}
\label{sec:conclusion}

We developed a representation-aware framework for analyzing and improving closed-surface parameterization, focusing on genus-0 surfaces and spherical parameterization. For FLASH, the unmodified south-pole Beltrami correction was shown to coincide exactly with source-mesh harmonic extension under consistent piecewise-linear assembly. On a fixed-anchor branch, this identity yields a closed-form factorization and an explicit sensitivity Jacobian that separates chart transition, harmonic propagation, and inverse stereographic projection. Together with the finite-triangle conditioning bound, the factorization provides a mathematical basis for the bounded affine planar preconditioner.

Across 30 genus-0 closed surfaces, the preconditioner reduced median polar upper-tail log-condition and chord-angle distortion by $10.17\%$ and $10.97\%$, respectively, with both measures improving on every mesh. The south-pole correction strengthened the raw planar gains, and the equivalent harmonic realization reproduced the original correction to numerical precision while providing a median correction-stage speedup of $1.63\times$. Downstream experiments showed fewer residual flips in SDEM and increasingly consistent improvements in fixed-template remeshing as template resolution increased. The cMCF and SDEM experiments further showed that terminal attenuation can coexist with substantial transient amplification, stopping-index changes, and loss of geometric validity. These results identify peak trajectory response, convergence behavior, and geometric validity as useful signals for future initialization control. 

Overall, our work supports representation selection as an algorithmic design variable in closed-surface parameterization. In the spherical setting studied here, it affects map quality, influences the numerical behavior of iterative parameterization, and can transfer measurable benefits to subsequent geometric computation.
\bibliographystyle{abbrv}
\bibliography{references}
\end{document}